\documentclass[reprint,floatfix, aps,prc, twocolumn,notitlepage,
nofootinbib,nobibnotes,
amsmath,amssymb,preprintnumbers]{revtex4-2}
\usepackage{adjustbox}
\usepackage{graphicx,xcolor}
\usepackage{physics}
\usepackage{enumitem}   
\usepackage{dcolumn}
\usepackage{hyperref}

\usepackage[normalem]{ulem}
\usepackage{mathtools}
\usepackage{bm}
\usepackage{subfigure} 
\usepackage{xcolor}
\usepackage{float}
\usepackage{placeins}
\usepackage{tabularx}
\usepackage{url}
\usepackage{latexsym}
\usepackage{epsf}
\usepackage{booktabs}
\usepackage{etoolbox}
\usepackage{cleveref}
\usepackage{siunitx}
\usepackage[version=4]{mhchem}

\makeatletter
\appto\abstract{%
	\let\latexlist\list
	\def\list{\edef\keeprightskip{\the\rightskip}\latexlist}%
	\patchcmd\latexlist{\ignorespaces}{\rightskip\keeprightskip\ignorespaces}{}{}%
}
\makeatother

\def\hybrid{SMASH-vHLLE hybrid}

\newcommand{\Pythia}{\textsc{Pythia}}

\begin{document}
	\title{Collective effects in O-O and Ne-Ne collisions at $\sqrt{s_{\mathrm{NN}}}$=\SI{5.36}{\tera\electronvolt} from a hybrid approach}
	\author{Lucas Constantin$^{1,2}$, Niklas G\"otz$^{1,2}$, Carl B. Rosenkvist$^{1,2,3}$ and Hannah Elfner$^{3,1,2,4}$}
	\affiliation{$^1$Institute for Theoretical Physics, Department of Physics, Goethe University Frankfurt, 60438 Frankfurt, Germany}
	\affiliation{$^2$Frankfurt Institute for Advanced Studies,  60438
		Frankfurt am Main, Germany}
	\affiliation{$^3$GSI Helmholtzzentrum f\"ur Schwerionenforschung,  64291
		Darmstadt, Germany}
	\affiliation{$^4$Helmholtz Research Academy Hesse for FAIR (HFHF), GSI Helmholtz Center,
		Campus Frankfurt,  60438 Frankfurt am Main, Germany}
	
	\date{\today}
	\begin{abstract}
		Many features of heavy-ion collisions are well described by hybrid approaches, where the droplet of strongly coupled quark gluon plasma (QGP) is modeled by hydrodynamics and the subsequent dilute stage is performed with a hadronic transport model. Conventionally, the formation of a QGP is well established in larger collision systems like lead and gold. However, hints of collectivity were found even in proton-proton collisions, raising the question where the onset of QGP formation lays. This study aims at making predictions for the light ions run at the CERN Large Hadron Collider in July 2025, in order to explore the applicability of hybrid approaches in smaller collision systems. We employ three different models, the \hybrid~approach, the pure hadronic cascade of SMASH and Angantyr to simulate O-O collisions at a center-of-mass energy of $\sqrt{s_{\mathrm{NN}}}$=\SI{5.36}{\tera\electronvolt}. This setup allows us to compare evolutions with and without a hydrodynamic description on an equal basis, while Angantyr serves as a baseline for no collective effects. 
	\end{abstract}
	
	\maketitle
	\section{Introduction}
	Heavy-ion collisions provide access to the QCD phase diagram over a broad range of temperatures and baryon chemical potentials. In particular, they allow us to study matter under conditions similar to those microseconds after the big bang~\cite{Elfner:2022iae}. In recent decades, a vast amount of evidence for the formation of a deconfined, fluid-like quark-gluon plasma has accumulated, based on the characteristic signatures such a phase leaves on the final state particle distributions~\cite{Gyulassy:2004vg,Muller2012}.
    
    One of these imprints is anisotropic flow, which emerges because the initially anisotropic collision region creates a pressure gradient that pushes the resulting particles outward anisotropically. The initial-state spatial anisotropies are quantified by the eccentricities $\varepsilon_n$, while the final-state momentum anisotropies are typically characterized by the Fourier coefficients $v_n$ of the azimuthal particle distribution
	\begin{equation}
		\label{eq:final_fourier_series}
		\frac{\mathrm{d}N}{\differential\phi} = \frac{1}{2 \pi} \left( 1 + 2 \sum_{n = 1}^{\infty} v_n \mathrm{cos}(n (\phi - \Psi_{n})) \right). 
	\end{equation}
	
	Hybrid approaches have proven particularly successful in reproducing the observed correlations between initial eccentricities and final-state anisotropic flow by modeling heavy-ion collisions using viscous hydrodynamics for the hot and dense stage followed by a transport approach for the hadronic rescatterings of the resulting hadrons~\cite{Hirano:2012kj, Petersen:2014yqa, Gotz:2025wnv, Nijs:2020roc}. However, such signs of collectivity have been observed even in proton-proton collisions—namely the observation of long-range near-side angular correlations~\cite{CMS:2010ifv} and nonzero anisotropic flow coefficients in $p$-Pb collisions~\cite{ALICE:2012eyl}—contradicting the prior belief that collectivity is exclusive to heavy-ion collisions. Recently, the ALICE Collaboration even observed partonic flow in $p$-$p$ collisions~\cite{ALICE:2024vzv}. 
    Another indication of QGP formation can be observed using the nuclear modification factor 
    \begin{equation}
		\label{eq:nuclear_modification_factor}
		R_{\text{AA}}(p_T) = \frac{1}{\langle N_{\mathrm{coll}} \rangle} \frac{\differential^2N_{\text{AA}}/\differential p_T \differential y}{\differential^2N_{\mathrm{pp}}/ \differential p_T \differential y} \Big\rvert_{y = 0}, 
	\end{equation}
    where one compares the particle spectra of heavy-ion collisions to those in $p$-$p$ collisions, normalized by the number of binary nucleon-nucleon collisions $\langle N_{\mathrm{coll}} \rangle$. Deviations from unity therefore signal the presence of modifications induced by the created medium \cite{ALICE:2018ekf}. One such modification is jet quenching, i.e. the energy loss of particles with high transverse momentum moving through the QGP, leading to an observed suppression of high $p_T$ particles in the nuclear modification factor~\cite{PHENIX:2001hpc}. The significant centrality dependence of this effect was one of the first indications for QGP formation~\cite{ALICE:2018vuu}. At lower $p_T$, the medium instead enhances particle yields due to radial flow, where the collective expansion of the QGP boosts particles towards higher $p_T$~\cite{Lv:2014vza}.
    
    In contrast to anisotropic flow, jet quenching has not been observed in $p$-$p$ or $p$-Pb collisions, raising questions about the onset of QGP formation \cite{Grosse-Oetringhaus:2024bwr}. In particular, is a hydrodynamical description suitable to capture the collective effects observed in small collision systems? This question motivates the study of intermediate small systems like O-O and Ne-Ne because they bridge the region between high multiplicity $p$-$p$ and low multiplicity Pb-Pb collisions~\cite{Brewer:2021kiv}. These systems offer another interesting feature: they exhibit $\alpha$-clustered nuclear structures, where nucleons tend to form He-nuclei to maximize their binding energy. These structures leave imprints on the initial density profiles that get translated to the final state by the QGP expansion and can help us study the connection between nuclear structure and final-state QGP signatures. Previous studies have demonstrated this effect comprehensively within hybrid approaches~\cite{Giacalone:2024ixe, Prasad:2024ahm}. The goal of this paper is to present the emergence of the aforementioned collective effects in comparison to a non-collective baseline, as well as a hadronic transport evolution that allows us to do an appropriate comparison to the hybrid approach. The light ions run at the CERN Large Hadron Collider (LHC) in July 2025 will provide the experimental input needed to advance our understanding of collectivity in small systems. The paper is organized as follows. We begin by describing the models used for the simulations, followed by a discussion of the initial state, including an assessment of the applicability of hydrodynamics. Next, we briefly outline the method for calculating the anisotropic flow coefficients. Finally, we present results for the final state. 
	
	\section{Model Description}\label{sec:model}
	This section briefly summarizes the most important aspects of the three models that are employed for the simulations. For a more detailed description, we refer to the respective references. 
	\subsection{SMASH-vHLLE-hybrid approach}\label{sec:hybrid}
	The SMASH-vHLLE hybrid approach~\cite{Schafer:2021csj, hybridurl} is based on the hadronic transport approach SMASH~\cite{SMASH:2016zqf, smashurl} and the 3+1-dimensional viscous hydrodynamics code vHLLE~\cite{Karpenko:2013wva, vhlleurl}. The initial stage of the collision is performed with the hadronic evolution of SMASH, which solves the relativistic Boltzmann equations numerically on a spacetime grid. At the high beam energies of interest for this study, the SMASH transport evolution consists of the initial state for the colliding nuclei based on Woods-Saxon sampling or external configurations for realistic profiles (see Sec. \ref{sec:ic} for more details). When the nucleons collide, {\Pythia}~\cite{Bierlich:2022pfr} is employed to model the inelastic interactions by producing color strings that fragment into hadrons. These hadrons are propagated in SMASH with a formation time during which their interactions are suppressed, and only afterwards do they re-interact with their full hadronic cross-sections.

	To construct the initial state for the hydrodynamical evolution, several approaches exist, including IP-Glasma~\cite{Schenke:2012wb} and the 3D resolved McDipper~\cite{Garcia-Montero:2023gex}, that are based on gluon saturation physics. However, for a more direct comparison, we will obtain the initial state from the hadronic evolution of SMASH. At a switching time $\tau_{\mathrm{sw}}$, a hypersurface of constant proper time is defined and every particle that crosses this hypersurface gets removed from the evolution. 
	To start the hydrodynamical evolution from this list of particles, a Gaussian smearing kernel, Lorentz-contracted in the longitudinal direction, is applied to distribute the conserved quantities over the grid cells. 
	The hydrodynamical evolution is governed by the conservation of energy and momentum, as well as the net-charge, net-baryon, and net-strangeness number. There is a temperature dependent shear and bulk viscosity implemented according to the parametrizations used in \cite{Gotz:2025wnv}. The equation of state is based on a chiral model fitted to lattice QCD data as described in \cite{Motornenko:2019arp}. 
	The system expands and cools down, until a switching energy density $\varepsilon_{\text{sw}}$ is reached. For every timestep, a hypersurface of constant energy density is constructed. Assuming a grand-canonical ensemble, each surface element is particlized individually in two steps: First, the number of particles of each species is sampled from a Poisson distribution. Second, the particles' momenta are sampled according to the Cooper-Frye formula.   
	The hadronic evolution of the resulting particles is again performed by SMASH. The free parameters of the model are inferred from experimental data via a Bayesian analysis that was performed in \cite{Gotz:2025wnv}. However, it should be kept in mind that collisions at lower energies were used and that the resulting temperature dependent shear viscosity over entropy $\eta/s$ is strictly decreasing with $T$, leading to zero $\eta/s$ in hot cells. To demonstrate the impact of this parameter, we also perform another simulation with constant $\eta/s$=0.08, which is expected to increase $\eta/s$ in most cells compared to the temperature dependent parameter set.

	\subsection{Angantyr}
Angantyr~\cite{Bierlich:2018xfw} is an extension of the {\Pythia}~\cite{Bierlich:2022pfr, pythiaurl} framework, originally developed for p–p collisions, to simulate full heavy-ion interactions. 
\\
The procedure begins by sampling the spatial positions of nucleons in the colliding nuclei with a Monte Carlo Glauber model that incorporates subnucleonic
fluctuations. From this configuration, the wounded nucleons are identified, while distinguishing between different types of nucleon–nucleon encounters. Both projectile and target nucleons may undergo primary as well as secondary interactions.
\\
Each nucleon–nucleon sub-collision is then modeled with the {\Pythia} minimum-bias p–p routine, which generates an inelastic partonic final state including multiple partonic interactions, parton showers, and beam remnants. 
\\
The resulting partonic systems are then combined and hadronized with the Lund string fragmentation model, yielding a complete hadron-level heavy-ion event.
\\
Since the sub-collisions are treated as independent, Angantyr does not generate collective effects and thus provides a useful baseline for flow analyses.
    
	\section{Initial State}\label{sec:ic}
	In this section, we will analyze the initial state obtained with the procedure described in Sec.~\ref{sec:hybrid}. In particular, we want to assess the applicability of hydrodynamics in the smaller collision system at hand as well as the impact of nuclear clustering on the initial spatial anisotropy.
	\subsection{Nuclear configurations}
	To sample configurations of large and spherical nuclei, a Woods-Saxon distribution 
	\begin{equation}
		\label{eq:Woods-Saxon-dist}
		\frac{\differential N}{\differential^3r} = \frac{\rho_0}{e^{\frac{r-r_0}{d}} + 1},
	\end{equation}
	is usually employed, with the nuclear saturation density $\rho_0$, such that the integral over the Woods-Saxon distribution gives the number of nucleons, the nuclear radius $r_0$ and the diffusiveness $d$. However, as mentioned before, in \ce{^{16}O} and \ce{^{16}Ne}, there are clustering and deformation effects present, that need to be taken into account~\cite{Freer:2017gip}. To describe these effects from first principles, Nuclear Lattice Effective Field Theory (NLEFT) provides an \textit{ab initio} framework for addressing this complicated nuclear many-body  problem~\cite{Lee:2008fa}. NLEFT configurations for \ce{^{16}O} and \ce{^{20} Ne} are taken from~\cite{Giacalone:2024luz} and only positive weighted configurations are used. As demonstrated there, the $\alpha$-clustered \ce{^{16}O} configuration assumes a tetrahedron shape, while the \ce{^{20} Ne} configuration, that has one $\alpha$-cluster more, assumes a bowling-pin shape. Before coming to the final state results of the aforementioned observables, we will take a look at two important properties of the SMASH initial conditions: the opacity and the eccentricity. 
    
	\subsection{Eccentricity}
\begin{figure}[h]
		\centering 
		\includegraphics[width=0.5\textwidth]{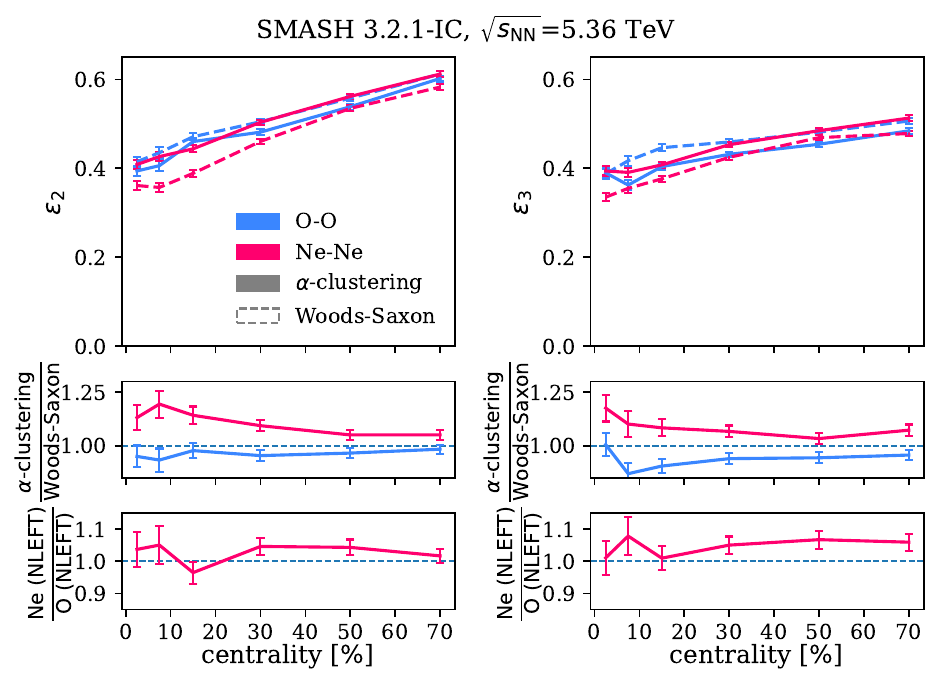}
		\caption{Comparison of the eccentricity in O-O and Ne-Ne collisions with Woods-Saxon and $\alpha$-clustered configurations in the SMASH initial conditions at $\sqrt{s_{\text{NN}}}$=\SI{5.36}{\tera\electronvolt}.}
		\label{fig:eccentricity}
	\end{figure}

    In this work, the eccentricity is computed from the energy density in the transverse plane using the formula
	\begin{equation}
		\label{eq:ecc_langlerangle}
		|\varepsilon_n| = \frac{ \sqrt{\langle r^n \mathrm{sin}(n\varphi) \rangle^2 + \langle  r^n \mathrm{cos}(n\varphi) \rangle ^2}}{\langle r^n \rangle},
	\end{equation}
	where $\langle \cdot \rangle$ denotes the energy density weighted average. 
	The eccentricity harmonics $\varepsilon_n$ are shown in Figure~\ref{fig:eccentricity} as a function of centrality. Expectedly, both $\varepsilon_2$ and $\varepsilon_3$ increase with centrality. However, in central collisions, the eccentricity is already very high compared to larger collision systems like Au-Au, hinting at the fact that event-by-event fluctuations dominate the creation of $\varepsilon_n$ rather than the almond shape of the collision region, or the geometric shape of the nuclei. Still, the $\alpha$-clustered configurations of \ce{^{16}O} and \ce{^{20} Ne} show opposite effects on the obtained $\varepsilon_n$, compared to the Woods-Saxon distribution. While the $\alpha$-clustered configuration in \ce{^{20} Ne} increases both $\varepsilon_2$ and $\varepsilon_3$, it decreases the eccentricity in \ce{^{16}O}. This result is likely very sensitive to the centrality selection as well as the applied smearing parameters. Nonetheless, the large increase in \ce{^{20}Ne} could stem from its bowling-pin shape, which could in principle lead to a large ellipticity if the nuclei collide body-to-body. On the other hand, a significant increase of $\varepsilon_3$ in the tetrahedron \ce{^{16}O} configurations is not observed. 
	
	\section{The Cumulant Method}\label{sec:cumulant_method}
	In general, non-vanishing anisotropic flow coefficients relate to the existence of correlations between the particles' azimuthal angles and the reaction plane. However, the reaction plane is not known in an experiment and the "true" reaction plane is not always aligned with the impact parameter axis due to the random positions of participants on an event-by-event basis. The cumulant method circumvents this problem by considering the correlations between tuples of particles. However, particles can be correlated because of many reasons, including resonance decays and Coulomb interactions between the outgoing hadrons. These correlations contribute to $v_n$ but are not a consequence of flow due to the QGP expansion and are therefore called "nonflow". This has to be considered especially in small collision systems, because they are of order $\frac{1}{N-1}$. In general, calculating cumulants of order k suppresses the contributions from correlations between up to $k$-1 particles and can therefore provide more accurate flow measurements by subtracting nonflow contributions~\cite{Borghini:2000sa}. Another possibility to reduce nonflow is by separating the event into subevents, as described in Sec. \ref{sec:subevents}. In the following, the procedure for calculating $v_n$ with the Q-cumulant method is briefly described. Therein, the Q or flow vectors are defined as 
	\begin{equation}
		Q_n = \sum_{i}^{M} = e^{in\phi_i},
		\label{eq:Q_vectors}
	\end{equation}
	with the azimuthal angles $\phi_i$ of the particles and the multiplicity of the event $M$. Then, the two-particle correlation in each event can be rewritten with the flow vectors as
	\begin{align}
		\label{eq:correlation_functions}
		\langle 2 \rangle = \langle e^{in(\phi_i - \phi_j)} \rangle = \frac{1}{\binom{M}{2}2!}\sum_{\substack{i,j=1 \\ i \ne j}}^{M}e^{in(\phi_i - \phi_j)} = \frac{|Q_n|^2 - M}{M(M-1)},
	\end{align}
	where in the numerator on the right-hand side one has to subtract the $M$ cases where the two selected particles are the same and the denominator is given by the number of particle pairs. The four-particle correlation can be calculated in a similar way:
	\begin{align}
		\langle 4 \rangle &=\nonumber \langle e^{in(\phi_i + \phi_j - \phi_k - \phi_l)} \rangle \\ &=\nonumber \frac{|Q_n|^4 + |Q_{2n}|^2  - 2 \text{Re}[Q_{2n} Q_n^* Q_n^*]}{M(M-1)(M-2)(M-3)} \\ 
		&\quad - 2 \frac{2 (M-2) Q_n Q_n^{*}- M(M-3) }{M(M-2)}. 
	\end{align}
	Then, the two- and four-particle cumulant can be defined by the weighted average over the events
	\begin{equation}
		\langle \langle 2 \rangle \rangle = \frac{\sum_{i} w_i \langle 2 \rangle_i}{\sum_{i} w_i} \quad \text{and} \quad  \langle \langle 4 \rangle \rangle = \frac{\sum_{i} w_i \langle 4 \rangle_i}{\sum_{i} w_i}, 
	\end{equation}
	where the weights are chosen to be the number of unordered pairs or quadruples, depending on the cumulant order, meaning
	\begin{align}
		\label{eq:weights}
		w_{\langle 2 \rangle,i} &= M(M-1) \\ w_{\langle 4 \rangle, i} &= M(M-1)(M-2)(M-3). 
	\end{align}
	This approach of selecting event weights was shown in~\cite{Bilandzic:2012wva} to minimize the statistical spread between events. From here, the two- and four-particle cumulants as well as the flow coefficients can be computed according to 
	\begin{align}
		c_n\{2\} &= \langle \langle 2 \rangle \rangle\nonumber \\ c_n\{4\} &= \langle \langle 4 \rangle \rangle - \langle \langle 2 \rangle \rangle^2
	\end{align}
	and
	\begin{align}
		\label{eq:multi-particle_flow}
		v_n\{2\} &= \sqrt{c_n\{2\}} \nonumber \\ v_n\{4\} &= \sqrt[4]{-c_n\{4\}}.
	\end{align}
	Further explanations for the formulas above can be found in~\cite{Bilandzic:2012wva} and~\cite{Bilandzic:2010jr}.
	
	\subsection{Subevent methods}\label{sec:subevents}
\begin{figure}[h]
		\centering 
		\includegraphics[width=0.25\textwidth]{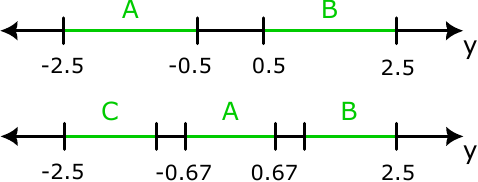}
		\caption[Rapidity ranges for the subevent methods]{Momentum rapidity ranges for the subevent methods. An approach similar to that in~\cite{Jia:2017hbm} is used.}
		\label{fig:subevents}
	\end{figure}

	Separating each event into multiple subevents with a gap in rapidity between them can reduce short-ranged nonflow correlations by not considering particle pairs that might have come from the same process. These methods are especially of interest if computing higher order cumulants is not feasible due to a lack of statistics, which is the case in smaller collision systems. In this work, the events are separated into two and three subevents based on the rapidity ranges shown in figure~\ref{fig:subevents}.
	The two-particle correlation from two subevents is given by
	\begin{equation}
		\label{eq:2_p_corr_subevents}
		\langle 2 \rangle_{a|b} = \frac{Q_{n,a} Q_{n,b}^{*}}{M_a M_b}, 
	\end{equation}
	where the $Q_{n,a/b}$ are the Q-vectors of the subevents and $M_{a/b}$ their multiplicities~\cite{Prasad:2024ahm}. For the four-particle cumulant, one can in principle construct a two, three or four subevent method. However, a three subevent method based on~\cite{Jia:2017hbm} is used for k. Therein, the four-particle cumulant is obtained using the formulas 
	\begin{equation}
		\label{eq:3sub_event_method}
		\langle 4 \rangle_{a,a|b,c} = \frac{(Q_{n,a}^2 - Q_{2n,a}) Q_{n,b}^* Q_{n,c}^*}{M_a (M_a - 1) M_b M_c}
	\end{equation}
	and 
	\begin{equation}
		\label{eq:3_subevent_cumulant}
		c_n^{a,a|b,c}\{4\} = \left\langle\!\left\langle 4 \right\rangle\!\right\rangle_{a,a|b,c} 
		- 2 \left\langle\!\left\langle 2 \right\rangle\!\right\rangle_{a|b} 
		\left\langle\!\left\langle 2 \right\rangle\!\right\rangle_{a|c}.
	\end{equation}

	\section{Final-State Results}
	\subsection{Centrality selection}
\begin{figure}[h]
		\centering 
		\includegraphics[width=0.5\textwidth]{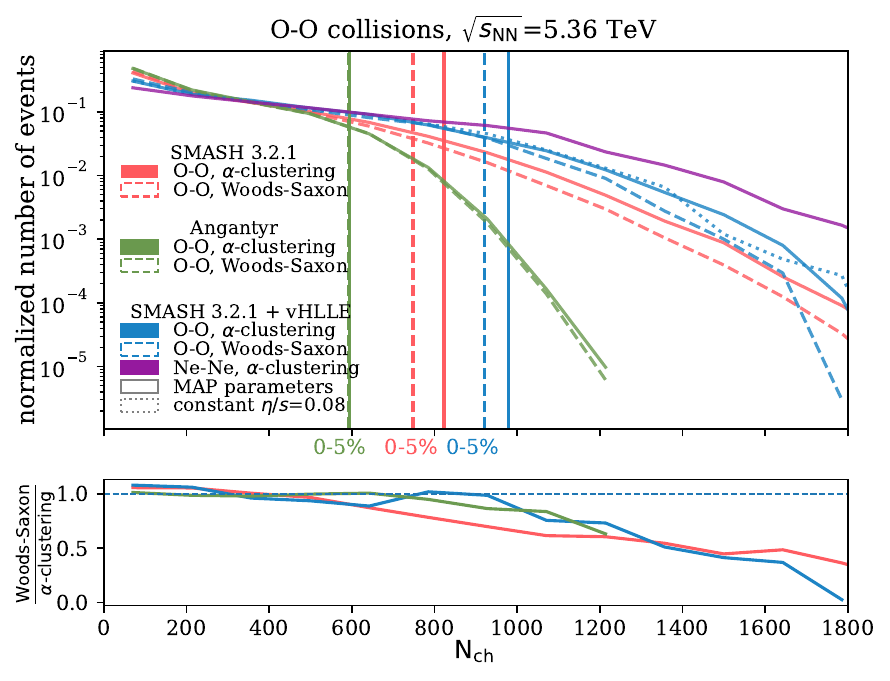}
		\caption{Multiplicity distribution in O-O collisions at $\sqrt{s_{\text{NN}}}$=\SI{5.36}{\tera\electronvolt} from SMASH transport (red), the SMASH-vHLLE hybrid model (blue) and Angantyr (green). The solid lines represent the $\alpha$-clustered configuration and the vertical lines indicate the lower bound of the 0-5\% centrality class.}
		\label{fig:multiplicity_dist}
	\end{figure}

	Before we can look at the collective effects, it is important to first look at the number of particles in the final states of the three models, because this affects the flow observables as well as the centrality selection. Figure~\ref{fig:multiplicity_dist} shows the charged multiplicity distribution of the three models. The SMASH-vHLLE hybrid approach produces higher multiplicity events than the pure transport and Angantyr. The final state charged multiplicity is an indicator for the entropy produced in the collision. From this perspective, the obtained results are sensible, because the hydrodynamical evolution describes a medium that is strongly coupled and where viscous effects continuously produce entropy, depending on the shear viscosity~\cite{Song:2008si}. In the SMASH transport on the other hand, the total interaction rate is lower, but entropy is still generated by the hadronic rescatterings and string excitations happening throughout the collision. Following the same logic, Angantyr produces much fewer particles than the other two models, since it does not perform hadronic rescatterings of the resulting hadrons. Furthermore, the $\alpha$-clustered configurations produce a marginally higher number of particles than the Woods-Saxon distribution in the hybrid and transport model, while the Angantyr results are almost unaffected by the nuclear geometry. We will therefore base the centrality selection on the charged particle multiplicity in the rapidity range $-2.5<y<2.5$, and select the the percentiles individually for each model and nuclear configuration. This results in different centrality bins for different configurations, as shown by the vertical lines in Figure~\ref{fig:multiplicity_dist}.
	
	\subsection{Nuclear modification factor}
\begin{figure*}[t]
		\centering 
		\includegraphics[width=0.8\textwidth]{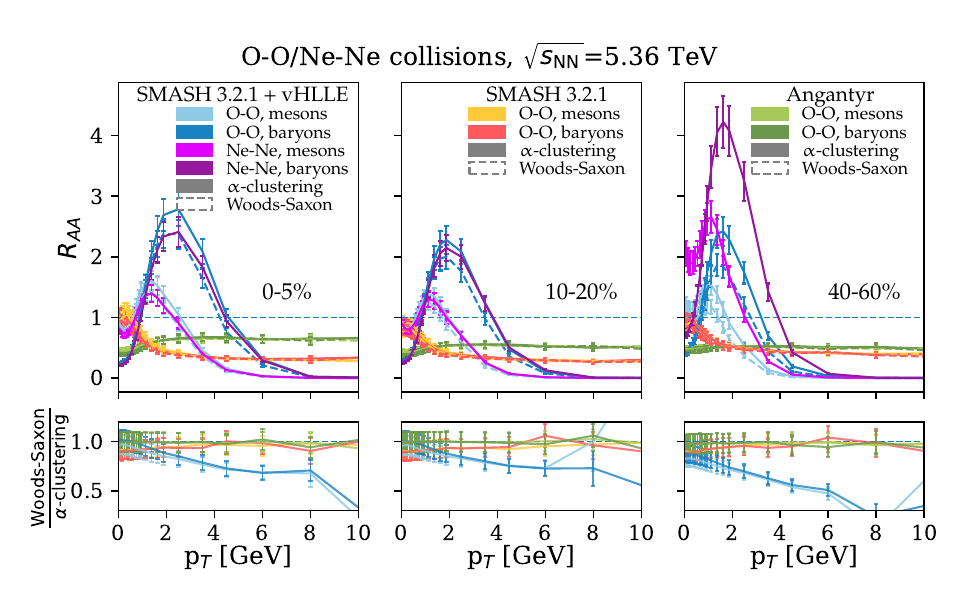}
		\caption{Nuclear modification factor $R_{\mathrm{AA}}$ in O-O and Ne-Ne collisions from the different models. }
		\label{fig:R_AA}
	\end{figure*}

	The nuclear modification factor compares the transverse momentum distributions of particles in nucleus-nucleus collisions to those in properly scaled $p$-$p$ collisions, in order to observe modifications due to the created medium. In reality, there are many effects present that influence the nuclear modification factor, including jet quenching, which causes a suppression at high $p_T$ and radial flow, pushing particles towards higher $p_T$ due to the QGP expansion~\cite{Lv:2014vza}. However, the SMASH-vHLLE hybrid approach currently does not include a treatment for extremely fast particles; rather, every particle enters the hydrodynamic evolution. Therefore, we will focus on the modifications at low $p_T$. Empirically, the expected $p_T$-spectrum in $p$-$p$ collisions follows an exponential decrease at low $p_T$ (below 2 GeV) and at high $p_T$ perturbative QCD predicts a power-law scaling, as described by the Hagedorn function~\cite{Saraswat:2017kpg}. In heavy-ion collisions, where one expects a thermalized QGP medium, the particle production can be described by the thermal model. Therein, the  production of hadrons is a statistical process in the grand-canonical ensemble, where the particles fill the phase-space according to a Boltzmann distribution~\cite{Braun-Munzinger:1994ewq}. The obtained $p_T$-spectra can be found in Appendix \ref{sec:p_T-spectry}. Finally, the number of nucleon-nucleon collisions $N_{\mathrm{coll}}$ is calculated from a Glauber model with a Woods-Saxon distribution. Figure~\ref{fig:R_AA} presents the nuclear modification factor for different centrality classes. The hybrid model shows a big enhancement in an intermediate $p_T$ region between 2 and 5 GeV and $R_{\mathrm{AA}}$ is clearly larger for baryons than for mesons. This is a clear sign of the particles being pushed from lower $p_T$ to higher $p_T$ by the hydrodynamic expansion of the QGP, because the heavier baryons gain more $p_T$ compared to the lighter mesons, leading to a bigger enhancement. In general, these results are qualitatively consistent with the expected result for the thermal divided by vacuum spectra. Furthermore, only the hybrid approach exhibits a significant difference between the two nuclear configurations, where the enhancement at intermediate $p_T$ is even larger for the $\alpha$-clustered configuration. This could hint at the medium being more dense in that case, resulting in a stronger radial flow expansion. The increase of $R_{\mathrm{AA}}$ with centrality is likely due to the number of binary nucleon collisions decreasing in the more peripheral collisions. 
	
	Second, the SMASH transport model shows an enhancement in $R_{\mathrm{AA}}$ at low $p_T$, and a constant suppression towards higher $p_T$, that decreases with centrality. 
	This can be attributed to the production of particles due to hadronic rescattering as well as the stopping of fast particles by the hadronic medium created in the collision. The subtle difference between baryons and mesons in SMASH is likely due the production of excess pions from resonance decays during the evolution. 
	On the other hand, the nuclear modification factor in Angantyr is approximately constant. 
	The fact that $R_{\mathrm{AA}}$ from Angantyr is below 1 at all $p_T$ hints at the fact, that it is not properly normalized by the number of binary collisions. A possible explanation for this is that secondary nucleon-nucleon interactions produce fewer particles, leading to a suppression when comparing to $\langle N_{\mathrm{coll}} \rangle$ $p$-$p$ collisions. The same reasoning could apply for the SMASH transport results. 
	In the following chapter, the anisotropic flow results will provide further insights into the collectivity of the models, in particular, the absence of collectivity in the pure hadronic transport of SMASH. 
	\subsection{Anisotropic flow}
\begin{figure}[h]
		\centering 
		\includegraphics[width=0.5\textwidth]{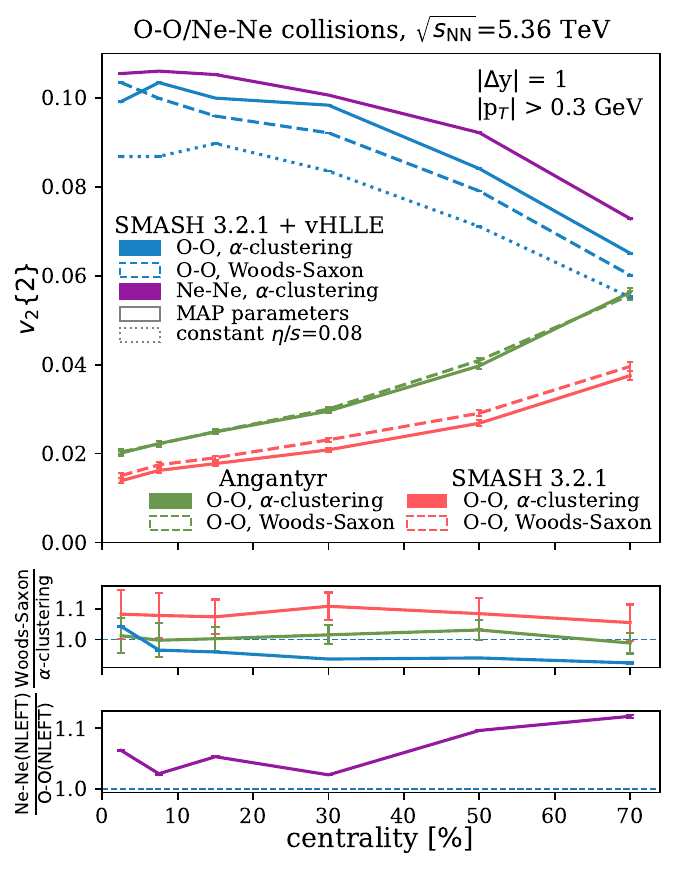}
		\caption{Elliptic flow $v_2\{2\}$ from the two-particle cumulant method with two subevents with a rapidity gap of $|\Delta y| = 1$ in O-O collisions at $\sqrt{s_{\text{NN}}}$=\SI{5.36}{\tera\electronvolt}.}
		\label{fig:integrated_v2}
	\end{figure}

	Now, we can finally look at the results obtained with the procedures outlined in Sec.~\ref{sec:cumulant_method}. Figure~\ref{fig:integrated_v2} shows the elliptic flow results from the two-particle cumulant method. One can see that $v_2\{2\}$ from the SMASH-vHLLE hybrid is the largest at central to mid-central collisions, but decreases for less central collisions. 
	
	Reflecting back on Figure~\ref{fig:eccentricity}, this could be a result from the fact that even the most central O-O collisions exhibit a high ellipticity compared to larger systems due to higher event-by-event fluctuations. The decrease in more peripheral collisions can be explained by the decreasing system size. As expected, a constant shear viscosity over entropy ratio significantly decreases the obtained $v_2\{2\}$, while the centrality dependence stays the same.
	
	The $\alpha$-clustered configurations of O and Ne both increase the anisotropic flow compared to the Woods-Saxon distribution. This is a result of the different densities of the created mediums, as shown in Figure~\ref{fig:R_AA}, and the difference in the nuclear shape, which becomes apparent from the enhancement of $v_2$ in the bowling-pin shaped Ne compared to the tetrahedron shaped O. 
	
	On the other hand, the hadronic transport SMASH and Angantyr show the opposite trend, where the flow increases in more peripheral collisions, with Angantyr even exceeding the flow obtained from the hybrid model in peripheral collisions. This indicates that neither approaches generate collective effects with the settings used in this work. Since Angantyr does not contain any collective behavior (no string shoving and ropes), one would naively expect it to have the smallest $v_2$ of the three models. However, the SMASH transport model actually has the smallest elliptic flow, while still matching the qualitative centrality dependence of Angantyr. This can be explained by the scaling of nonflow contributions with roughly $\frac{1}{N-1}$ and the fact that Angantyr produces much less particles in the final state, as shown in Figure~\ref{fig:multiplicity_dist}. Angantyr shows no difference between between nuclear configurations, while the small difference in SMASH is due to the different centrality selections for the two configurations and the fact that the $\alpha$-clustered configuration creates on average events with more final state particles, resulting in less nonflow contributions.
	
\begin{figure}[h]
		\centering 
		\includegraphics[width=0.5\textwidth]{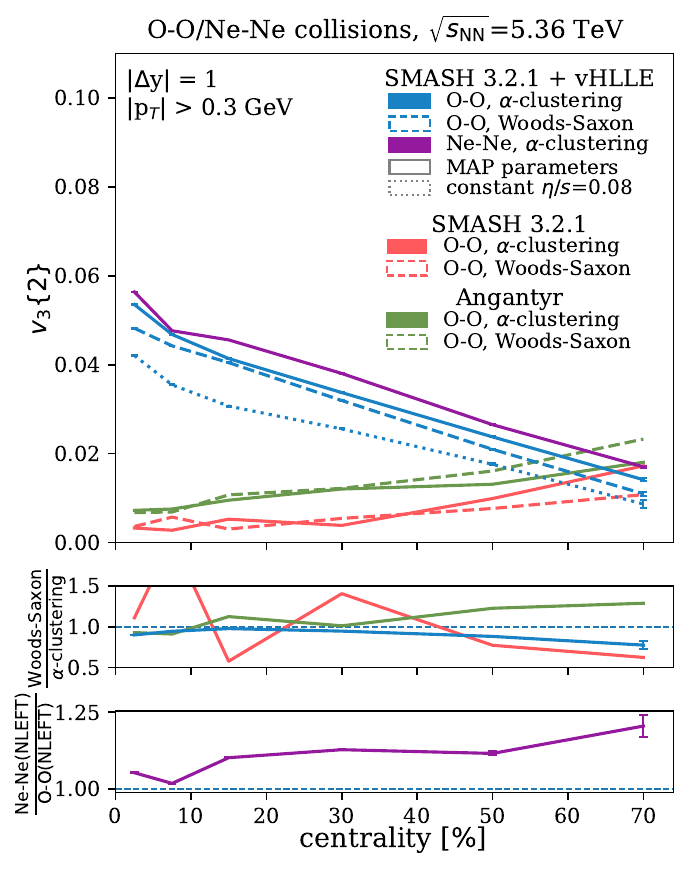}
		\caption{Triangular flow $v_3\{2\}$ from the two-particle cumulant method with two subevents with a rapidity gap of $|\Delta y| = 1$ in O-O collisions at $\sqrt{s_{\text{NN}}}$=\SI{5.36}{\tera\electronvolt}.}
		\label{fig:integrated_v3}
	\end{figure}

	Regarding the triangular flow in Figure~\ref{fig:integrated_v3}, one observes a similar pattern, where $v_3$ from the hybrid model is once again highest in central collisions but decreases for less central events, in contrast to the other models, which show the opposite trend. The same reasoning can be applied here, that explains the decrease in the hybrid by the decreasing system size and the increase in the other models by the decrease of final state particles, enhancing the nonflow contributions. 
	
	\subsubsection{Anisotropic flow fluctuations}
	When computing the two- and four-particle correlations to estimate the anisotropic flow, we actually obtain estimates for the squared and quadrupled flow coefficients $\langle v_n^2 \rangle$ and $\langle v_n^4 \rangle$ respectively. This is a consequence of Eq.~\eqref{eq:multi-particle_flow} and means that the calculated flow harmonics will be systematically biased by flow fluctuations, according to the general formula 
	\begin{equation}
		\label{eq:flow_fluctuations}
		\sigma^2_{v_n} =\langle v_n^2 \rangle - \langle v_n \rangle^2.  
	\end{equation}
	In~\cite{Ollitrault:2009ie} it was shown that the flow fluctuations increase the estimate from the two-particle cumulant method and decrease the estimate for the four-particle cumulant method: 
	\begin{equation}
		\label{eq:impact_flow_fluctuations}
		v_n^2\{2\} = \langle v_n \rangle^2 + \sigma^2_{v_n} \qquad v_n^2\{4\} \approx \langle v_n \rangle^2 - \sigma^2_{v_n}
	\end{equation}
	Ignoring nonflow, which affects $v_n\{2\}$ more than $v_n\{4\}$, one can easily derive the equations for the unbiased flow estimate $\langle v_n \rangle$ as well as the flow fluctuations $\sigma_{v_n}$
	\begin{align}
		\langle v_n \rangle &= \sqrt{\frac{ v_n^2\{2\} + v_n^2\{4\}}{2}} \nonumber \\ \sigma_{v_n} &= \sqrt{\frac{ v_n^2\{2\} - v_n^2\{4\}}{2}}.
	\end{align}
	\begin{figure}[h]
		\centering 
		\includegraphics[width=0.5\textwidth]{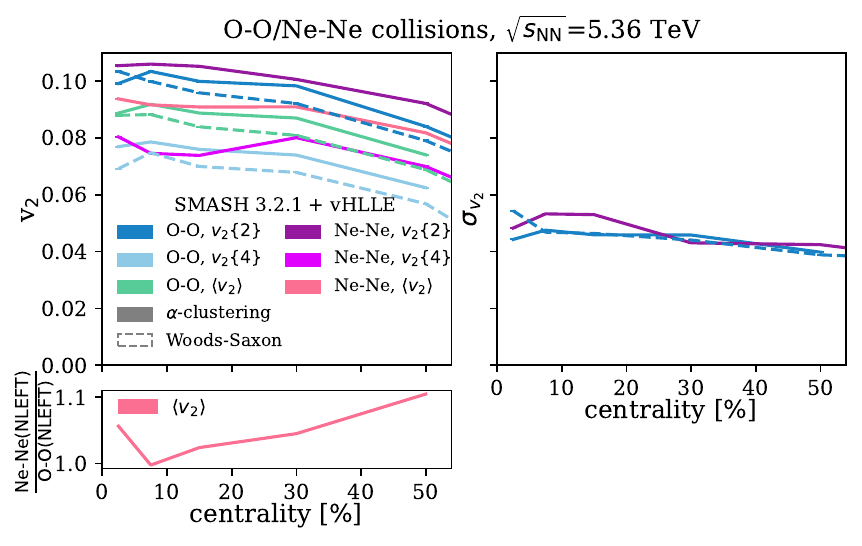}
		\caption{$v_2\{2\}$, $v_2\{4\}$ and $\langle v_2 \rangle$ (left) and the anisotropic flow fluctuations $\sigma_{v_n}$ (right) in O-O collisions from the SMASH-vHLLE hybrid model. The results where calculated using the 3 subevents method.}
		\label{fig:v2_4_finalO}
	\end{figure}
	\begin{figure}[h]
		\centering
		\includegraphics[width=0.5\textwidth]{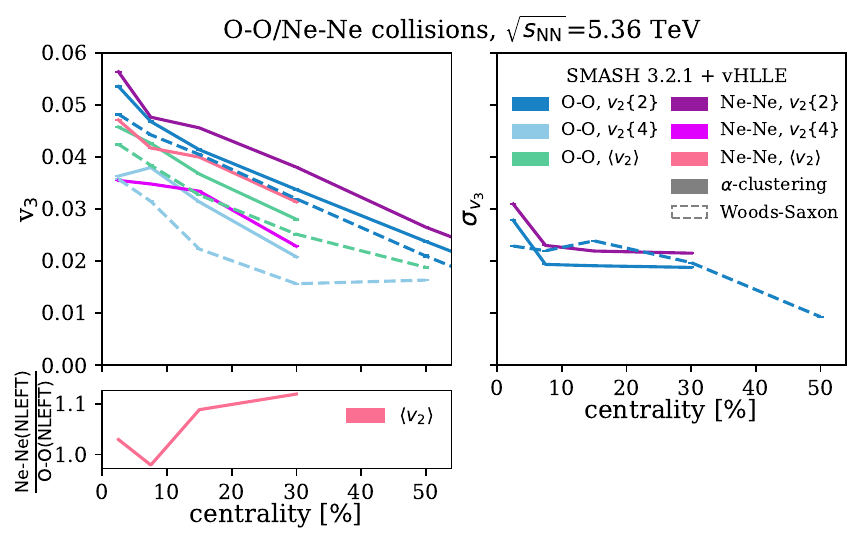}
		\caption{$v_3\{2\}$, $v_3\{4\}$ and $\langle v_3 \rangle$ (left) and the anisotropic flow fluctuations $\sigma_{v_n}$ (right) in O-O collisions from the SMASH-vHLLE hybrid model. The results where calculated using the 3 subevents method.}
		\label{fig:v3_4_finalO}
	\end{figure}
	The four-particle cumulants of the three models can be found in Appendix \ref{sec:4-particle_cumulants}. The fact that they are close to zero in both the SMASH transport and Angantyr confirms the hypothesis that the flow signal from these models stem from nonflow contributions. The hybrid approach exhibits the correct sign for the cumulants, allowing us to compute the four-particle elliptic flow. 
	Figure~\ref{fig:v2_4_finalO} shows $v_n\{4\}$ from the three subevents method and the resulting unbiased flow coefficient $\langle v_2 \rangle$ as well as the anisotropic flow fluctuations from the hybrid model. Interestingly, $v_n\{4\}$ shows the same trend as $v_n\{2\}$, at lower values. The flow fluctuations $\sigma_{v_n}$ therefore are almost constant with respect to the centrality. Possible sources for these fluctuations include fluctuations from the positions of the nucleons and multiplicity fluctuations~\cite{Bilandzic:2012wva}. The former would be expected to be higher in more peripheral collisions and the latter depends on the centrality selection. Naturally, the 0-5\% centrality class is the one with the largest range of multiplicity. However, other effects might play into this.
	\subsubsection{Differential flow}
\begin{figure}[h]
		\centering
		\hspace*{-1.7em}
		\includegraphics[width=0.53\textwidth]{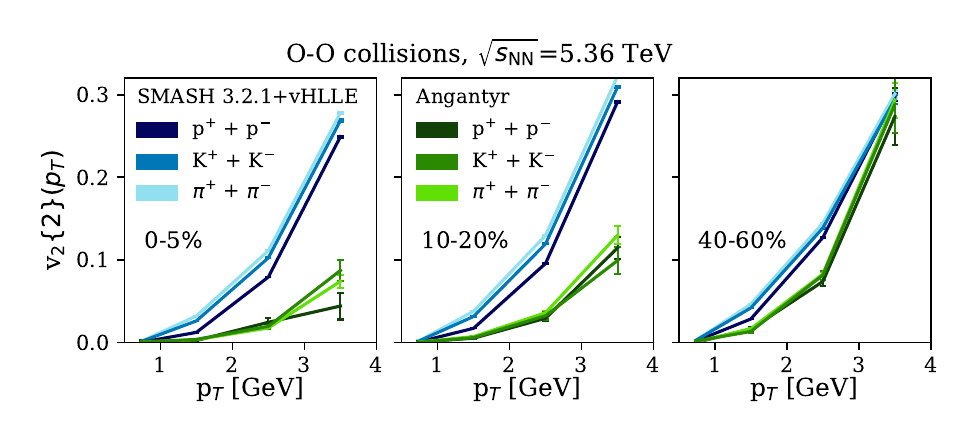}
		\caption{Differential flow $v_2\{2\}(p_T)$ from the two-particle cumulant method with two subevents with a rapidity gap of $|\Delta y| = 1$ in O-O collisions at $\sqrt{s_{\text{NN}}}$=\SI{5.36}{\tera\electronvolt}.}
		\label{fig:differential_v2}
	\end{figure}

	Lastly, the resulting differential $v_n\{2\}$ as a function of $p_T$ is shown in Figure~\ref{fig:differential_v2}. In the results from the hybrid model, the expected trend is visible, where $v_2$ rises with $p_T$ due to the expansion of the QGP and a mass ordering of $v_2$ from heavy to light particles can be observed, due to the heavier particles attaining more $p_T$ than lighter ones. Towards higher $p_T$, one would experimentally expect a baryon-meson splitting, due to the quarks coalescing together, where the individual $v_2$ of the partons adds up, so the baryons end up with higher anisotropic flow. However, there is no description of individual partons in the SMASH-vHLLE hybrid and also no coalescence happening. Therefore and since there is not a lot of hadronic rescattering that could also introduce the splitting, no significant baryon-meson splitting is observed in the results. The results from the Angantyr model demonstrate that nonflow contributions might play a larger role at higher $p_T$ and more peripheral collisions. 
	
	\section{Conclusions and outlook}
	In this work, O-O and Ne-Ne collisions at a center of mass energy of  $\sqrt{s_{\text{NN}}}$=\SI{5.36}{\tera\electronvolt} were simulated using three different modeling approaches, including the hadronic transport evolution SMASH, the SMASH-vHLLE hybrid approach and Angantyr, with the goal to analyze the emergence of collective effects in hybrid approaches.
	
	The anisotropic flow coefficients were calculated from multiparticle cumulant methods with subevents, in order to reduce nonflow contributions. We observe clear differences between the collective and non-collective models, namely opposite trends in $v_n$ and mass ordering patterns in the differential flow results. 

	Although the absolute value of the flow results from the hybrid model are highly parameter dependent, the quantitative behavior is not, resulting in a characteristic decrease with centrality. We find that the SMASH transport and Angantyr are dominated by nonflow, highlighting the importance of higher order cumulants and nonflow reduction measures when analyzing flow signals from small collision systems, which is crucial for the interpretation of experimental data from the LHC. 
	
	The nuclear modification factor from the hybrid approach shows the expected results for particle spectra from the thermal model divided by vacuum spectra. The observed mass ordering of $R_{\mathrm{AA}}$ is a clear sign of radial flow due to the QGP expansion. In addition, a clear difference emerges between the two nuclear configurations, with the $\alpha$-clustered configuration increasing $R_{\mathrm{AA}}$. The pure hadronic evolution of SMASH shows only a small enhancement below 1 GeV from particles being decelerated by the medium, while Angantyr shows almost no modification at all. 
	In conclusion, if there is a fluid-like QGP forming in O-O collisions, we expect a big enhancement of $v_2$ and $v_3$ in central collisions and that a significant difference between nuclear structure is only observed if collective dynamics are present in the system. We also highlights the importance of properly treating nonflow contributions and flow fluctuations when simulating small collision systems. The experimental data presented in \cite{ATLAS:2025nnt} will be essential for improving our understanding of collectivity from small to large collisions. 
	\begin{acknowledgments}
		This work was funded by the Deutsche Forschungsgemeinschaft (DFG, German Research Foundation), Project No. 315477589 – TRR 211. Computational resources were provided by the GreenCube at GSI. The authors want to thank Renan Góes-Hirayama for helpful discussions.
	\end{acknowledgments}
	
	\appendix
	\section{Applicability of Hydrodynamics}
\begin{figure}[h]
		\centering
		\includegraphics[width=0.4\textwidth]{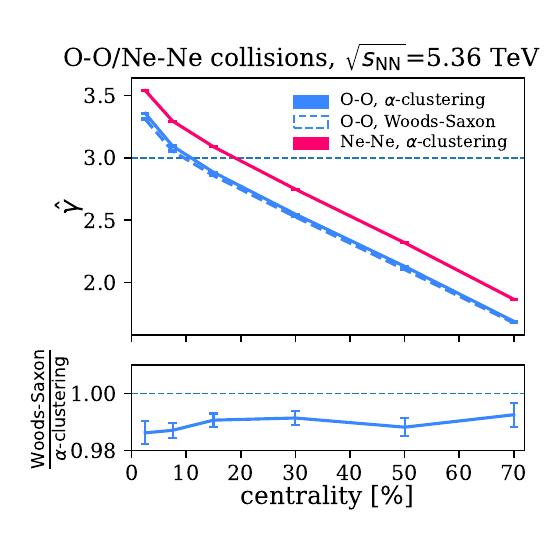}
		\caption[Opacity as a function of centrality]{Opacity as a function of centrality in O-O and Ne-Ne collisions in the SMASH initial conditions at the switching time from \cite{Gotz:2025wnv}.} 
	\label{fig:opacity}
	\end{figure}

	The switch between a transport description to hydrodynamics relies on the assumption of local thermal equilibrium. However, this condition is not necessarily true on an event-by-event basis~\cite{Oliinychenko:2015lva,Inghirami:2022afu}, especially in small collision systems~\cite{Noronha:2024dtq}. There are multiple ways to quantify how close the system is to local thermal equilibrium, such as the Knudsen number or the inverse Reynolds number. Here, we opt for the easily accessible opacity 
	\begin{equation}
		\label{eq:opacity}
		\hat{\gamma} = \left(5 \eta/s \right)^{-1}\left( \frac{1}{a \pi} R \frac{\differential E_{T}}{\differential\eta_s}\right)^{\frac{1}{4}},
	\end{equation}
	which is a measure for the interaction rate in the system and depends on the shear viscosity over entropy $\eta/s$, the transverse radius $R$ and the energy deposition per rapidity $\frac{\mathrm{d}E_{T}}{\mathrm{d}\eta_s}$~\cite{Kurkela:2019kip}. The parameter $a$ is the proportionality constant in the conformal equation of state $\varepsilon = a T^4$ and is set to $42.25 \frac{\pi^2}{30}$ according to \cite{Ambrus:2022koq}.  
	To calculate the transverse radius from the SMASH initial state, the transverse area that exceeds an energy density of 0.5~GeV is calculated and assumed to be circular. In~\cite{Ambrus:2022qya} it was shown that a hydrodynamic description remains accurate above $\hat{\gamma} \approx 3-4$ compared to kinetic theory, which is applicable under less restrictive conditions. 
	Figure~\ref{fig:opacity} presents results for the opacity. We find that $\hat{\gamma}$ exceeds 3 in the top 10\% most central events and decreases linearly towards more peripheral collisions. This suggests that hydrodynamics provides a reliable description in central O-O collisions, while its accuracy becomes worse in more peripheral events. This emphasizes why light-ion collisions are interesting, because observables that are sensitive to this potential breakdown of the applicability of hydrodynamics can help us understand the origin of collective effects in small systems. To see where our theoretical predictions differ from experimental data, we run the hydrodynamical description in every centrality class, despite the limited applicability.
	
	\section{$p_T$-spectra}\label{sec:p_T-spectry}
\begin{figure}[h]
		\centering
		\includegraphics[width=0.5\textwidth]{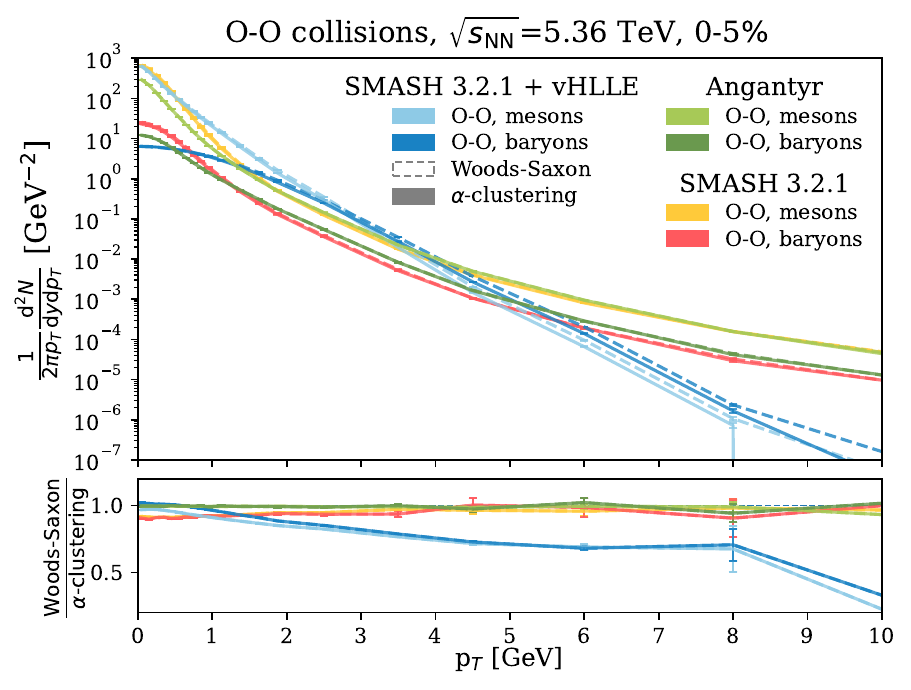}
		\caption[$p_T$-Spectra of baryons and mesons in O-O collisions]{$p_T$-spectra of baryons and mesons from O-O collisions at $\sqrt{s_{\text{NN}}}$= \SI{5.36}{\tera\electronvolt} in the 0-5\% centrality class.}
		\label{fig:pT_spectra_OO}	
	\end{figure}

	Figure~\ref{fig:pT_spectra_OO} shows the $p_T$-spectra that were used to calculate the nuclear modification factor. The transverse momentum distribution from the hybrid approach shows a big enhancement of baryons and mesons between 2 and 5 GeV, and a suppression of low and high $p_T$ particles in O-O collisions, compared to the other two models. With respect to the hybrid results, the spectra from the SMASH transport and Angantyr look qualitatively similar. Still, we observe that SMASH produces more soft baryons and mesons compared to Angantyr. With respect to Figure~\ref{fig:multiplicity_dist}, we note that most of the excess particles in the final state of the SMASH transport compared to Angantyr are low $p_T$ mesons. 
	
	\section{four-particle cumulants}\label{sec:4-particle_cumulants}
\begin{figure}[h]
		\centering 
		\includegraphics[width=0.5\textwidth]{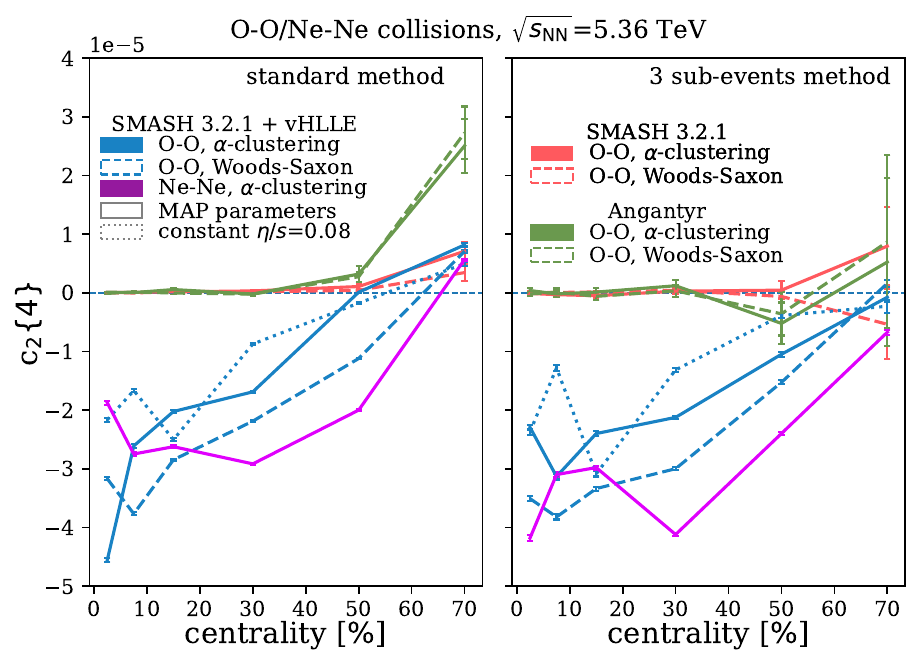}
		\caption{four-particle cumulant $c_2\{4\}$ from the standard method and the three-subevent-method with a rapidity gap of $|\Delta y| = 1$ (right) in O-O collisions at $\sqrt{s_{\text{NN}}}$=\SI{5.36}{\tera\electronvolt}.}
		\label{fig:c2_4}
	\end{figure}

\begin{figure}[h]
		\centering 
		\includegraphics[width=0.5\textwidth]{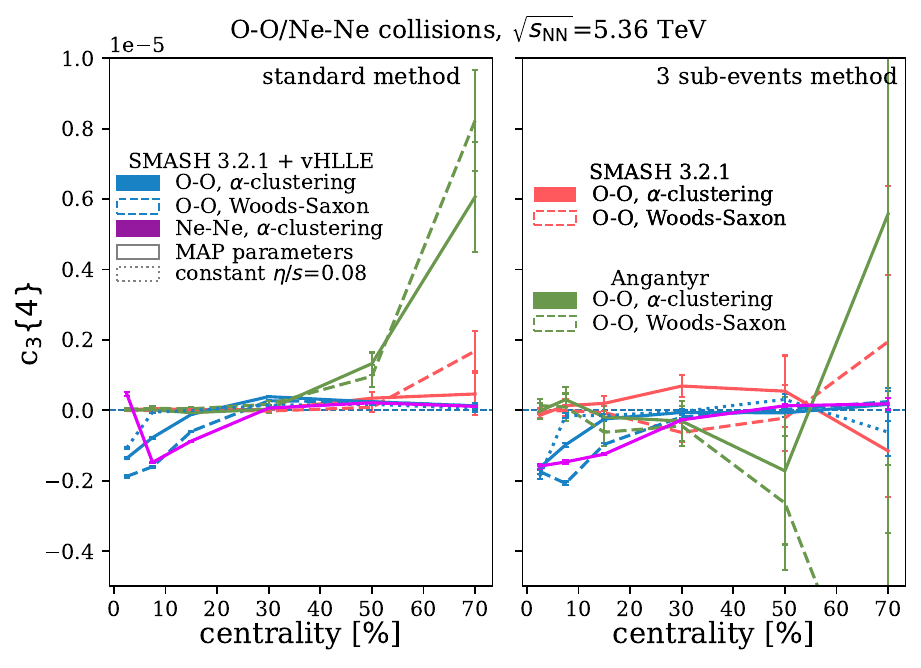}
		\caption{four-particle cumulant $c_3\{4\}$ from the standard method and the three-subevent-method with a rapidity gap of $|\Delta y| = 1$ (right) in O-O collisions at $\sqrt{s_{\text{NN}}}$=\SI{5.36}{\tera\electronvolt}.}
		\label{fig:c3_4}
	\end{figure}

	As Figures~\ref{fig:c2_4} and \ref{fig:c3_4} show, the $c_n\{4\}$ are mostly close to zero for SMASH and Angantyr, confirming the hypothesis that their flow results are dominated by nonflow effects. As expected, only the hybrid model exhibits large and negative valued $c_n\{4\}$, allowing us to calculate the four-particle flow coefficients up to the 40-60\% centrality class. 
	
	\section{Nuclear modification factor from $\pi^0$}
\begin{figure}[H]
		\centering 
		\includegraphics[width=0.45\textwidth]{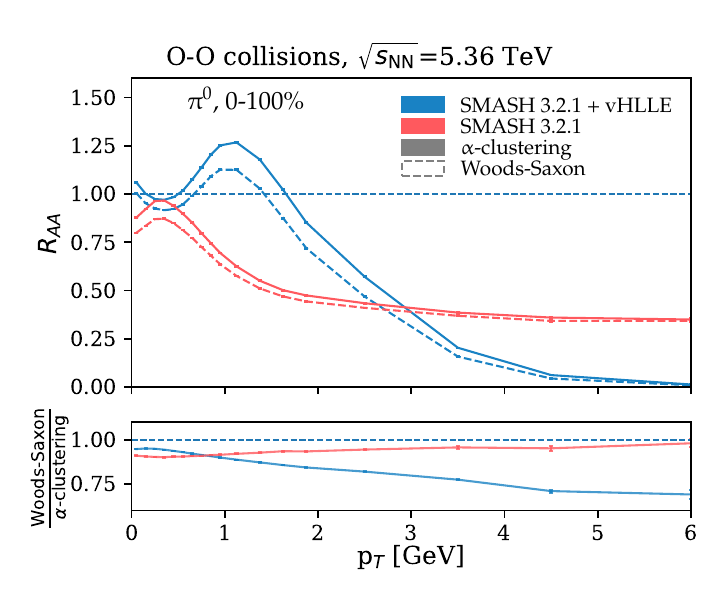}
		\caption{Nuclear modification factor $R_{\mathrm{AA}}$ in O-O collisions from $\pi^0$ with 0-100\% centrality.}
		\label{fig:R_OO_pi0}
	\end{figure}

	For better comparison to experimental data we present the nuclear modification factor from $\pi^0$ in Figure~\ref{fig:R_OO_pi0}. The general features described above still hold, however the peak structure in the purely hadronic transport and the hybrid approach are closer together, highlighting the importance of a good resolution in the low $p_T$ region. 
	
	\section{Software Versions}
	The software components used in this work are publicly available and summarized in Table~\ref{tab:model-versions}. 
	\begin{table}[H]
		\centering
		\caption{Versions of the physics models and tools used in this work.}
		\begin{tabular}{lc}
			\toprule
			\textbf{Physics software} & \textbf{Version / Tag} \\
			\midrule
			Pythia8 \cite{pythiaurl}          & 8.315 \\
			SMASH \cite{smashurl}             & 3.2.1 \\
			SMASH-vHLLE-Hybrid \cite{hybridurl}      & 2.0 \\
			vHLLE \cite{vhlleurl}                   & vhlle-smash-hybrid-1 \\
			SMASH hadron sampler     & same as SMASH \\
			\bottomrule
		\end{tabular}
		\label{tab:model-versions}
	\end{table}
	\FloatBarrier
	\bibliography{2025_oxygen_bib}

\begin{thebibliography}{48}%
\makeatletter
\providecommand \@ifxundefined [1]{%
 \@ifx{#1\undefined}
}%
\providecommand \@ifnum [1]{%
 \ifnum #1\expandafter \@firstoftwo
 \else \expandafter \@secondoftwo
 \fi
}%
\providecommand \@ifx [1]{%
 \ifx #1\expandafter \@firstoftwo
 \else \expandafter \@secondoftwo
 \fi
}%
\providecommand \natexlab [1]{#1}%
\providecommand \enquote  [1]{``#1''}%
\providecommand \bibnamefont  [1]{#1}%
\providecommand \bibfnamefont [1]{#1}%
\providecommand \citenamefont [1]{#1}%
\providecommand \href@noop [0]{\@secondoftwo}%
\providecommand \href [0]{\begingroup \@sanitize@url \@href}%
\providecommand \@href[1]{\@@startlink{#1}\@@href}%
\providecommand \@@href[1]{\endgroup#1\@@endlink}%
\providecommand \@sanitize@url [0]{\catcode `\\12\catcode `\$12\catcode
  `\&12\catcode `\#12\catcode `\^12\catcode `\_12\catcode `\%12\relax}%
\providecommand \@@startlink[1]{}%
\providecommand \@@endlink[0]{}%
\providecommand \url  [0]{\begingroup\@sanitize@url \@url }%
\providecommand \@url [1]{\endgroup\@href {#1}{\urlprefix }}%
\providecommand \urlprefix  [0]{URL }%
\providecommand \Eprint [0]{\href }%
\providecommand \doibase [0]{https://doi.org/}%
\providecommand \selectlanguage [0]{\@gobble}%
\providecommand \bibinfo  [0]{\@secondoftwo}%
\providecommand \bibfield  [0]{\@secondoftwo}%
\providecommand \translation [1]{[#1]}%
\providecommand \BibitemOpen [0]{}%
\providecommand \bibitemStop [0]{}%
\providecommand \bibitemNoStop [0]{.\EOS\space}%
\providecommand \EOS [0]{\spacefactor3000\relax}%
\providecommand \BibitemShut  [1]{\csname bibitem#1\endcsname}%
\let\auto@bib@innerbib\@empty
\bibitem [{\citenamefont {Elfner}\ and\ \citenamefont
  {M\"uller}(2023)}]{Elfner:2022iae}%
  \BibitemOpen
  \bibfield  {author} {\bibinfo {author} {\bibfnamefont {H.}~\bibnamefont
  {Elfner}}\ and\ \bibinfo {author} {\bibfnamefont {B.}~\bibnamefont
  {M\"uller}},\ }\bibfield  {title} {\bibinfo {title} {{The exploration of hot
  and dense nuclear matter: introduction to relativistic heavy-ion physics}},\
  }\href {https://doi.org/10.1088/1361-6471/ace824} {\bibfield  {journal}
  {\bibinfo  {journal} {J. Phys. G}\ }\textbf {\bibinfo {volume} {50}},\
  \bibinfo {pages} {103001} (\bibinfo {year} {2023})},\ \Eprint
  {https://arxiv.org/abs/2210.12056} {arXiv:2210.12056 [nucl-th]} \BibitemShut
  {NoStop}%
\bibitem [{\citenamefont {Gyulassy}(2004)}]{Gyulassy:2004vg}%
  \BibitemOpen
  \bibfield  {author} {\bibinfo {author} {\bibfnamefont {M.}~\bibnamefont
  {Gyulassy}},\ }\bibfield  {title} {\bibinfo {title} {{The QGP discovered at
  RHIC}},\ }in\ \href@noop {} {\emph {\bibinfo {booktitle} {{NATO Advanced
  Study Institute: Structure and Dynamics of Elementary Matter}}}}\ (\bibinfo
  {year} {2004})\ pp.\ \bibinfo {pages} {159--182},\ \Eprint
  {https://arxiv.org/abs/nucl-th/0403032} {arXiv:nucl-th/0403032} \BibitemShut
  {NoStop}%
\bibitem [{\citenamefont {Muller}\ \emph {et~al.}(2012)\citenamefont {Muller},
  \citenamefont {Schukraft},\ and\ \citenamefont {Wyslouch}}]{Muller2012}%
  \BibitemOpen
  \bibfield  {author} {\bibinfo {author} {\bibfnamefont {B.}~\bibnamefont
  {Muller}}, \bibinfo {author} {\bibfnamefont {J.}~\bibnamefont {Schukraft}},\
  and\ \bibinfo {author} {\bibfnamefont {B.}~\bibnamefont {Wyslouch}},\
  }\bibfield  {title} {\bibinfo {title} {{First Results from Pb+Pb collisions
  at the LHC}},\ }\href {https://doi.org/10.1146/annurev-nucl-102711-094910}
  {\bibfield  {journal} {\bibinfo  {journal} {Ann. Rev. Nucl. Part. Sci.}\
  }\textbf {\bibinfo {volume} {62}},\ \bibinfo {pages} {361} (\bibinfo {year}
  {2012})},\ \Eprint {https://arxiv.org/abs/1202.3233} {arXiv:1202.3233
  [hep-ex]} \BibitemShut {NoStop}%
\bibitem [{\citenamefont {Hirano}\ \emph {et~al.}(2013)\citenamefont {Hirano},
  \citenamefont {Huovinen}, \citenamefont {Murase},\ and\ \citenamefont
  {Nara}}]{Hirano:2012kj}%
  \BibitemOpen
  \bibfield  {author} {\bibinfo {author} {\bibfnamefont {T.}~\bibnamefont
  {Hirano}}, \bibinfo {author} {\bibfnamefont {P.}~\bibnamefont {Huovinen}},
  \bibinfo {author} {\bibfnamefont {K.}~\bibnamefont {Murase}},\ and\ \bibinfo
  {author} {\bibfnamefont {Y.}~\bibnamefont {Nara}},\ }\bibfield  {title}
  {\bibinfo {title} {{Integrated Dynamical Approach to Relativistic Heavy Ion
  Collisions}},\ }\href {https://doi.org/10.1016/j.ppnp.2013.02.002} {\bibfield
   {journal} {\bibinfo  {journal} {Prog. Part. Nucl. Phys.}\ }\textbf {\bibinfo
  {volume} {70}},\ \bibinfo {pages} {108} (\bibinfo {year} {2013})},\ \Eprint
  {https://arxiv.org/abs/1204.5814} {arXiv:1204.5814 [nucl-th]} \BibitemShut
  {NoStop}%
\bibitem [{\citenamefont {Petersen}(2014)}]{Petersen:2014yqa}%
  \BibitemOpen
  \bibfield  {author} {\bibinfo {author} {\bibfnamefont {H.}~\bibnamefont
  {Petersen}},\ }\bibfield  {title} {\bibinfo {title} {{Anisotropic flow in
  transport + hydrodynamics hybrid approaches}},\ }\href
  {https://doi.org/10.1088/0954-3899/41/12/124005} {\bibfield  {journal}
  {\bibinfo  {journal} {J. Phys. G}\ }\textbf {\bibinfo {volume} {41}},\
  \bibinfo {pages} {124005} (\bibinfo {year} {2014})},\ \Eprint
  {https://arxiv.org/abs/1404.1763} {arXiv:1404.1763 [nucl-th]} \BibitemShut
  {NoStop}%
\bibitem [{\citenamefont {G{\"o}tz}\ \emph {et~al.}(2025)\citenamefont
  {G{\"o}tz}, \citenamefont {Karpenko},\ and\ \citenamefont
  {Elfner}}]{Gotz:2025wnv}%
  \BibitemOpen
  \bibfield  {author} {\bibinfo {author} {\bibfnamefont {N.}~\bibnamefont
  {G{\"o}tz}}, \bibinfo {author} {\bibfnamefont {I.}~\bibnamefont {Karpenko}},\
  and\ \bibinfo {author} {\bibfnamefont {H.}~\bibnamefont {Elfner}},\
  }\bibfield  {title} {\bibinfo {title} {{Bayesian analysis of a (3+1)D hybrid
  approach with initial conditions from hadronic transport}},\ }\href
  {https://doi.org/10.1103/rzml-rjxz} {\bibfield  {journal} {\bibinfo
  {journal} {Phys. Rev. C}\ }\textbf {\bibinfo {volume} {112}},\ \bibinfo
  {pages} {014910} (\bibinfo {year} {2025})},\ \Eprint
  {https://arxiv.org/abs/2503.10181} {arXiv:2503.10181 [nucl-th]} \BibitemShut
  {NoStop}%
\bibitem [{\citenamefont {Nijs}\ \emph {et~al.}(2021)\citenamefont {Nijs},
  \citenamefont {van~der Schee}, \citenamefont {G{\"u}rsoy},\ and\
  \citenamefont {Snellings}}]{Nijs:2020roc}%
  \BibitemOpen
  \bibfield  {author} {\bibinfo {author} {\bibfnamefont {G.}~\bibnamefont
  {Nijs}}, \bibinfo {author} {\bibfnamefont {W.}~\bibnamefont {van~der Schee}},
  \bibinfo {author} {\bibfnamefont {U.}~\bibnamefont {G{\"u}rsoy}},\ and\
  \bibinfo {author} {\bibfnamefont {R.}~\bibnamefont {Snellings}},\ }\bibfield
  {title} {\bibinfo {title} {{Bayesian analysis of heavy ion collisions with
  the heavy ion computational framework Trajectum}},\ }\href
  {https://doi.org/10.1103/PhysRevC.103.054909} {\bibfield  {journal} {\bibinfo
   {journal} {Phys. Rev. C}\ }\textbf {\bibinfo {volume} {103}},\ \bibinfo
  {pages} {054909} (\bibinfo {year} {2021})},\ \Eprint
  {https://arxiv.org/abs/2010.15134} {arXiv:2010.15134 [nucl-th]} \BibitemShut
  {NoStop}%
\bibitem [{\citenamefont {Khachatryan}\ \emph {et~al.}(2010)\citenamefont
  {Khachatryan} \emph {et~al.}}]{CMS:2010ifv}%
  \BibitemOpen
  \bibfield  {author} {\bibinfo {author} {\bibfnamefont {V.}~\bibnamefont
  {Khachatryan}} \emph {et~al.} (\bibinfo {collaboration} {CMS}),\ }\bibfield
  {title} {\bibinfo {title} {{Observation of Long-Range Near-Side Angular
  Correlations in Proton-Proton Collisions at the LHC}},\ }\href
  {https://doi.org/10.1007/JHEP09(2010)091} {\bibfield  {journal} {\bibinfo
  {journal} {JHEP}\ }\textbf {\bibinfo {volume} {09}},\ \bibinfo {pages}
  {091}},\ \Eprint {https://arxiv.org/abs/1009.4122} {arXiv:1009.4122 [hep-ex]}
  \BibitemShut {NoStop}%
\bibitem [{\citenamefont {Abelev}\ \emph {et~al.}(2013)\citenamefont {Abelev}
  \emph {et~al.}}]{ALICE:2012eyl}%
  \BibitemOpen
  \bibfield  {author} {\bibinfo {author} {\bibfnamefont {B.}~\bibnamefont
  {Abelev}} \emph {et~al.} (\bibinfo {collaboration} {ALICE}),\ }\bibfield
  {title} {\bibinfo {title} {{Long-range angular correlations on the near and
  away side in $p$-Pb collisions at $\sqrt{s_{NN}}=5.02$ TeV}},\ }\href
  {https://doi.org/10.1016/j.physletb.2013.01.012} {\bibfield  {journal}
  {\bibinfo  {journal} {Phys. Lett. B}\ }\textbf {\bibinfo {volume} {719}},\
  \bibinfo {pages} {29} (\bibinfo {year} {2013})},\ \Eprint
  {https://arxiv.org/abs/1212.2001} {arXiv:1212.2001 [nucl-ex]} \BibitemShut
  {NoStop}%
\bibitem [{\citenamefont {Acharya}\ \emph {et~al.}(2024)\citenamefont {Acharya}
  \emph {et~al.}}]{ALICE:2024vzv}%
  \BibitemOpen
  \bibfield  {author} {\bibinfo {author} {\bibfnamefont {S.}~\bibnamefont
  {Acharya}} \emph {et~al.} (\bibinfo {collaboration} {ALICE}),\ }\bibfield
  {title} {\bibinfo {title} {{Observation of partonic flow in proton-proton and
  proton-nucleus collisions}},\ }\href@noop {} {\  (\bibinfo {year} {2024})},\
  \Eprint {https://arxiv.org/abs/2411.09323} {arXiv:2411.09323 [nucl-ex]}
  \BibitemShut {NoStop}%
\bibitem [{\citenamefont {Acharya}\ \emph {et~al.}(2019)\citenamefont {Acharya}
  \emph {et~al.}}]{ALICE:2018ekf}%
  \BibitemOpen
  \bibfield  {author} {\bibinfo {author} {\bibfnamefont {S.}~\bibnamefont
  {Acharya}} \emph {et~al.} (\bibinfo {collaboration} {ALICE}),\ }\bibfield
  {title} {\bibinfo {title} {{Analysis of the apparent nuclear modification in
  peripheral Pb{\textendash}Pb collisions at 5.02 TeV}},\ }\href
  {https://doi.org/10.1016/j.physletb.2019.04.047} {\bibfield  {journal}
  {\bibinfo  {journal} {Phys. Lett. B}\ }\textbf {\bibinfo {volume} {793}},\
  \bibinfo {pages} {420} (\bibinfo {year} {2019})},\ \Eprint
  {https://arxiv.org/abs/1805.05212} {arXiv:1805.05212 [nucl-ex]} \BibitemShut
  {NoStop}%
\bibitem [{\citenamefont {Adcox}\ \emph {et~al.}(2002)\citenamefont {Adcox}
  \emph {et~al.}}]{PHENIX:2001hpc}%
  \BibitemOpen
  \bibfield  {author} {\bibinfo {author} {\bibfnamefont {K.}~\bibnamefont
  {Adcox}} \emph {et~al.} (\bibinfo {collaboration} {PHENIX}),\ }\bibfield
  {title} {\bibinfo {title} {{Suppression of hadrons with large transverse
  momentum in central Au+Au collisions at $\sqrt{s_{NN}}$ = 130-GeV}},\ }\href
  {https://doi.org/10.1103/PhysRevLett.88.022301} {\bibfield  {journal}
  {\bibinfo  {journal} {Phys. Rev. Lett.}\ }\textbf {\bibinfo {volume} {88}},\
  \bibinfo {pages} {022301} (\bibinfo {year} {2002})},\ \Eprint
  {https://arxiv.org/abs/nucl-ex/0109003} {arXiv:nucl-ex/0109003} \BibitemShut
  {NoStop}%
\bibitem [{\citenamefont {Acharya}\ \emph {et~al.}(2018)\citenamefont {Acharya}
  \emph {et~al.}}]{ALICE:2018vuu}%
  \BibitemOpen
  \bibfield  {author} {\bibinfo {author} {\bibfnamefont {S.}~\bibnamefont
  {Acharya}} \emph {et~al.} (\bibinfo {collaboration} {ALICE}),\ }\bibfield
  {title} {\bibinfo {title} {{Transverse momentum spectra and nuclear
  modification factors of charged particles in pp, p-Pb and Pb-Pb collisions at
  the LHC}},\ }\href {https://doi.org/10.1007/JHEP11(2018)013} {\bibfield
  {journal} {\bibinfo  {journal} {JHEP}\ }\textbf {\bibinfo {volume} {11}},\
  \bibinfo {pages} {013}},\ \Eprint {https://arxiv.org/abs/1802.09145}
  {arXiv:1802.09145 [nucl-ex]} \BibitemShut {NoStop}%
\bibitem [{\citenamefont {Lv}\ \emph {et~al.}(2014)\citenamefont {Lv},
  \citenamefont {Ma}, \citenamefont {Zhang}, \citenamefont {Chen},\ and\
  \citenamefont {Fang}}]{Lv:2014vza}%
  \BibitemOpen
  \bibfield  {author} {\bibinfo {author} {\bibfnamefont {M.}~\bibnamefont
  {Lv}}, \bibinfo {author} {\bibfnamefont {Y.~G.}\ \bibnamefont {Ma}}, \bibinfo
  {author} {\bibfnamefont {G.~Q.}\ \bibnamefont {Zhang}}, \bibinfo {author}
  {\bibfnamefont {J.~H.}\ \bibnamefont {Chen}},\ and\ \bibinfo {author}
  {\bibfnamefont {D.~Q.}\ \bibnamefont {Fang}},\ }\bibfield  {title} {\bibinfo
  {title} {{Nuclear modification factor in intermediate-energy heavy-ion
  collisions}},\ }\href {https://doi.org/10.1016/j.physletb.2014.04.025}
  {\bibfield  {journal} {\bibinfo  {journal} {Phys. Lett. B}\ }\textbf
  {\bibinfo {volume} {733}},\ \bibinfo {pages} {105} (\bibinfo {year}
  {2014})},\ \Eprint {https://arxiv.org/abs/1404.4394} {arXiv:1404.4394
  [nucl-th]} \BibitemShut {NoStop}%
\bibitem [{\citenamefont {Grosse-Oetringhaus}\ and\ \citenamefont
  {Wiedemann}(2024)}]{Grosse-Oetringhaus:2024bwr}%
  \BibitemOpen
  \bibfield  {author} {\bibinfo {author} {\bibfnamefont {J.~F.}\ \bibnamefont
  {Grosse-Oetringhaus}}\ and\ \bibinfo {author} {\bibfnamefont {U.~A.}\
  \bibnamefont {Wiedemann}},\ }\bibfield  {title} {\bibinfo {title} {{A Decade
  of Collectivity in Small Systems}},\ }\href@noop {} {\  (\bibinfo {year}
  {2024})},\ \Eprint {https://arxiv.org/abs/2407.07484} {arXiv:2407.07484
  [hep-ex]} \BibitemShut {NoStop}%
\bibitem [{\citenamefont {Brewer}\ \emph {et~al.}(2021)\citenamefont {Brewer},
  \citenamefont {Mazeliauskas},\ and\ \citenamefont {van~der
  Schee}}]{Brewer:2021kiv}%
  \BibitemOpen
  \bibfield  {author} {\bibinfo {author} {\bibfnamefont {J.}~\bibnamefont
  {Brewer}}, \bibinfo {author} {\bibfnamefont {A.}~\bibnamefont
  {Mazeliauskas}},\ and\ \bibinfo {author} {\bibfnamefont {W.}~\bibnamefont
  {van~der Schee}},\ }\bibfield  {title} {\bibinfo {title} {{Opportunities of
  OO and $p$O collisions at the LHC}},\ }in\ \href@noop {} {\emph {\bibinfo
  {booktitle} {{Opportunities of OO and pO collisions at the LHC}}}}\ (\bibinfo
  {year} {2021})\ \Eprint {https://arxiv.org/abs/2103.01939} {arXiv:2103.01939
  [hep-ph]} \BibitemShut {NoStop}%
\bibitem [{\citenamefont {Giacalone}\ \emph
  {et~al.}(2025{\natexlab{a}})\citenamefont {Giacalone} \emph
  {et~al.}}]{Giacalone:2024ixe}%
  \BibitemOpen
  \bibfield  {author} {\bibinfo {author} {\bibfnamefont {G.}~\bibnamefont
  {Giacalone}} \emph {et~al.},\ }\bibfield  {title} {\bibinfo {title}
  {{Anisotropic Flow in Fixed-Target Pb208+Ne20 Collisions as a Probe of
  Quark-Gluon Plasma}},\ }\href
  {https://doi.org/10.1103/PhysRevLett.134.082301} {\bibfield  {journal}
  {\bibinfo  {journal} {Phys. Rev. Lett.}\ }\textbf {\bibinfo {volume} {134}},\
  \bibinfo {pages} {082301} (\bibinfo {year} {2025}{\natexlab{a}})},\ \Eprint
  {https://arxiv.org/abs/2405.20210} {arXiv:2405.20210 [nucl-th]} \BibitemShut
  {NoStop}%
\bibitem [{\citenamefont {Prasad}\ \emph {et~al.}(2025)\citenamefont {Prasad},
  \citenamefont {Mallick}, \citenamefont {Sahoo},\ and\ \citenamefont
  {Barnaf{\"o}ldi}}]{Prasad:2024ahm}%
  \BibitemOpen
  \bibfield  {author} {\bibinfo {author} {\bibfnamefont {S.}~\bibnamefont
  {Prasad}}, \bibinfo {author} {\bibfnamefont {N.}~\bibnamefont {Mallick}},
  \bibinfo {author} {\bibfnamefont {R.}~\bibnamefont {Sahoo}},\ and\ \bibinfo
  {author} {\bibfnamefont {G.~G.}\ \bibnamefont {Barnaf{\"o}ldi}},\ }\bibfield
  {title} {\bibinfo {title} {{Anisotropic flow fluctuation as a possible
  signature of clustered nuclear geometry in O{\textendash}O collisions at the
  Large Hadron Collider}},\ }\href
  {https://doi.org/10.1016/j.physletb.2024.139145} {\bibfield  {journal}
  {\bibinfo  {journal} {Phys. Lett. B}\ }\textbf {\bibinfo {volume} {860}},\
  \bibinfo {pages} {139145} (\bibinfo {year} {2025})},\ \Eprint
  {https://arxiv.org/abs/2407.15065} {arXiv:2407.15065 [nucl-th]} \BibitemShut
  {NoStop}%
\bibitem [{\citenamefont {Sch\"afer}\ \emph {et~al.}(2022)\citenamefont
  {Sch\"afer}, \citenamefont {Karpenko}, \citenamefont {Wu}, \citenamefont
  {Hammelmann},\ and\ \citenamefont {Elfner}}]{Schafer:2021csj}%
  \BibitemOpen
  \bibfield  {author} {\bibinfo {author} {\bibfnamefont {A.}~\bibnamefont
  {Sch\"afer}}, \bibinfo {author} {\bibfnamefont {I.}~\bibnamefont {Karpenko}},
  \bibinfo {author} {\bibfnamefont {X.-Y.}\ \bibnamefont {Wu}}, \bibinfo
  {author} {\bibfnamefont {J.}~\bibnamefont {Hammelmann}},\ and\ \bibinfo
  {author} {\bibfnamefont {H.}~\bibnamefont {Elfner}} (\bibinfo {collaboration}
  {SMASH}),\ }\bibfield  {title} {\bibinfo {title} {{Particle production in a
  hybrid approach for a beam energy scan of Au+Au/Pb+Pb collisions between
  $\sqrt{s_\textrm{NN}}$ = 4.3 GeV and $\sqrt{s_\textrm{NN}}$ = 200.0 GeV}},\
  }\href {https://doi.org/10.1140/epja/s10050-022-00872-x} {\bibfield
  {journal} {\bibinfo  {journal} {Eur. Phys. J. A}\ }\textbf {\bibinfo {volume}
  {58}},\ \bibinfo {pages} {230} (\bibinfo {year} {2022})},\ \Eprint
  {https://arxiv.org/abs/2112.08724} {arXiv:2112.08724 [hep-ph]} \BibitemShut
  {NoStop}%
\bibitem [{hyb()}]{hybridurl}%
  \BibitemOpen
  \href {https://github.com/smash-transport/smash-vhlle-hybrid} {}\bibinfo
  {howpublished}
  {\url{https://github.com/smash-transport/smash-vhlle-hybrid}}\BibitemShut
  {NoStop}%
\bibitem [{\citenamefont {Weil}\ \emph {et~al.}(2016)\citenamefont {Weil} \emph
  {et~al.}}]{SMASH:2016zqf}%
  \BibitemOpen
  \bibfield  {author} {\bibinfo {author} {\bibfnamefont {J.}~\bibnamefont
  {Weil}} \emph {et~al.} (\bibinfo {collaboration} {SMASH}),\ }\bibfield
  {title} {\bibinfo {title} {{Particle production and equilibrium properties
  within a new hadron transport approach for heavy-ion collisions}},\ }\href
  {https://doi.org/10.1103/PhysRevC.94.054905} {\bibfield  {journal} {\bibinfo
  {journal} {Phys. Rev. C}\ }\textbf {\bibinfo {volume} {94}},\ \bibinfo
  {pages} {054905} (\bibinfo {year} {2016})},\ \Eprint
  {https://arxiv.org/abs/1606.06642} {arXiv:1606.06642 [nucl-th]} \BibitemShut
  {NoStop}%
\bibitem [{sma()}]{smashurl}%
  \BibitemOpen
  \href {https://smash-transport.github.io/} {}\bibinfo {howpublished}
  {\url{https://smash-transport.github.io/}}\BibitemShut {NoStop}%
\bibitem [{\citenamefont {Karpenko}\ \emph {et~al.}(2014)\citenamefont
  {Karpenko}, \citenamefont {Huovinen},\ and\ \citenamefont
  {Bleicher}}]{Karpenko:2013wva}%
  \BibitemOpen
  \bibfield  {author} {\bibinfo {author} {\bibfnamefont {I.}~\bibnamefont
  {Karpenko}}, \bibinfo {author} {\bibfnamefont {P.}~\bibnamefont {Huovinen}},\
  and\ \bibinfo {author} {\bibfnamefont {M.}~\bibnamefont {Bleicher}},\
  }\bibfield  {title} {\bibinfo {title} {{A 3+1 dimensional viscous
  hydrodynamic code for relativistic heavy ion collisions}},\ }\href
  {https://doi.org/10.1016/j.cpc.2014.07.010} {\bibfield  {journal} {\bibinfo
  {journal} {Comput. Phys. Commun.}\ }\textbf {\bibinfo {volume} {185}},\
  \bibinfo {pages} {3016} (\bibinfo {year} {2014})},\ \Eprint
  {https://arxiv.org/abs/1312.4160} {arXiv:1312.4160 [nucl-th]} \BibitemShut
  {NoStop}%
\bibitem [{vhl()}]{vhlleurl}%
  \BibitemOpen
  \href {https://github.com/yukarpenko/vhlle} {}\bibinfo {howpublished}
  {\url{https://github.com/yukarpenko/vhlle}}\BibitemShut {NoStop}%
\bibitem [{\citenamefont {Bierlich}\ \emph {et~al.}(2022)\citenamefont
  {Bierlich} \emph {et~al.}}]{Bierlich:2022pfr}%
  \BibitemOpen
  \bibfield  {author} {\bibinfo {author} {\bibfnamefont {C.}~\bibnamefont
  {Bierlich}} \emph {et~al.},\ }\bibfield  {title} {\bibinfo {title} {{A
  comprehensive guide to the physics and usage of PYTHIA 8.3}},\ }\href
  {https://doi.org/10.21468/SciPostPhysCodeb.8} {\bibfield  {journal} {\bibinfo
   {journal} {SciPost Phys. Codeb.}\ }\textbf {\bibinfo {volume} {2022}},\
  \bibinfo {pages} {8} (\bibinfo {year} {2022})},\ \Eprint
  {https://arxiv.org/abs/2203.11601} {arXiv:2203.11601 [hep-ph]} \BibitemShut
  {NoStop}%
\bibitem [{\citenamefont {Schenke}\ \emph {et~al.}(2012)\citenamefont
  {Schenke}, \citenamefont {Tribedy},\ and\ \citenamefont
  {Venugopalan}}]{Schenke:2012wb}%
  \BibitemOpen
  \bibfield  {author} {\bibinfo {author} {\bibfnamefont {B.}~\bibnamefont
  {Schenke}}, \bibinfo {author} {\bibfnamefont {P.}~\bibnamefont {Tribedy}},\
  and\ \bibinfo {author} {\bibfnamefont {R.}~\bibnamefont {Venugopalan}},\
  }\bibfield  {title} {\bibinfo {title} {{Fluctuating Glasma initial conditions
  and flow in heavy ion collisions}},\ }\href
  {https://doi.org/10.1103/PhysRevLett.108.252301} {\bibfield  {journal}
  {\bibinfo  {journal} {Phys. Rev. Lett.}\ }\textbf {\bibinfo {volume} {108}},\
  \bibinfo {pages} {252301} (\bibinfo {year} {2012})},\ \Eprint
  {https://arxiv.org/abs/1202.6646} {arXiv:1202.6646 [nucl-th]} \BibitemShut
  {NoStop}%
\bibitem [{\citenamefont {Garcia-Montero}\ \emph {et~al.}(2024)\citenamefont
  {Garcia-Montero}, \citenamefont {Elfner},\ and\ \citenamefont
  {Schlichting}}]{Garcia-Montero:2023gex}%
  \BibitemOpen
  \bibfield  {author} {\bibinfo {author} {\bibfnamefont {O.}~\bibnamefont
  {Garcia-Montero}}, \bibinfo {author} {\bibfnamefont {H.}~\bibnamefont
  {Elfner}},\ and\ \bibinfo {author} {\bibfnamefont {S.}~\bibnamefont
  {Schlichting}},\ }\bibfield  {title} {\bibinfo {title} {{McDIPPER: A novel
  saturation-based 3+1D initial-state model for heavy ion collisions}},\ }\href
  {https://doi.org/10.1103/PhysRevC.109.044916} {\bibfield  {journal} {\bibinfo
   {journal} {Phys. Rev. C}\ }\textbf {\bibinfo {volume} {109}},\ \bibinfo
  {pages} {044916} (\bibinfo {year} {2024})},\ \Eprint
  {https://arxiv.org/abs/2308.11713} {arXiv:2308.11713 [hep-ph]} \BibitemShut
  {NoStop}%
\bibitem [{\citenamefont {Motornenko}\ \emph {et~al.}(2020)\citenamefont
  {Motornenko}, \citenamefont {Steinheimer}, \citenamefont {Vovchenko},
  \citenamefont {Schramm},\ and\ \citenamefont
  {Stoecker}}]{Motornenko:2019arp}%
  \BibitemOpen
  \bibfield  {author} {\bibinfo {author} {\bibfnamefont {A.}~\bibnamefont
  {Motornenko}}, \bibinfo {author} {\bibfnamefont {J.}~\bibnamefont
  {Steinheimer}}, \bibinfo {author} {\bibfnamefont {V.}~\bibnamefont
  {Vovchenko}}, \bibinfo {author} {\bibfnamefont {S.}~\bibnamefont {Schramm}},\
  and\ \bibinfo {author} {\bibfnamefont {H.}~\bibnamefont {Stoecker}},\
  }\bibfield  {title} {\bibinfo {title} {{Equation of state for hot QCD and
  compact stars from a mean field approach}},\ }\href
  {https://doi.org/10.1103/PhysRevC.101.034904} {\bibfield  {journal} {\bibinfo
   {journal} {Phys. Rev. C}\ }\textbf {\bibinfo {volume} {101}},\ \bibinfo
  {pages} {034904} (\bibinfo {year} {2020})},\ \Eprint
  {https://arxiv.org/abs/1905.00866} {arXiv:1905.00866 [hep-ph]} \BibitemShut
  {NoStop}%
\bibitem [{\citenamefont {Bierlich}\ \emph {et~al.}(2018)\citenamefont
  {Bierlich}, \citenamefont {Gustafson}, \citenamefont {L{\"o}nnblad},\ and\
  \citenamefont {Shah}}]{Bierlich:2018xfw}%
  \BibitemOpen
  \bibfield  {author} {\bibinfo {author} {\bibfnamefont {C.}~\bibnamefont
  {Bierlich}}, \bibinfo {author} {\bibfnamefont {G.}~\bibnamefont {Gustafson}},
  \bibinfo {author} {\bibfnamefont {L.}~\bibnamefont {L{\"o}nnblad}},\ and\
  \bibinfo {author} {\bibfnamefont {H.}~\bibnamefont {Shah}},\ }\bibfield
  {title} {\bibinfo {title} {{The Angantyr model for Heavy-Ion Collisions in
  PYTHIA8}},\ }\href {https://doi.org/10.1007/JHEP10(2018)134} {\bibfield
  {journal} {\bibinfo  {journal} {JHEP}\ }\textbf {\bibinfo {volume} {10}},\
  \bibinfo {pages} {134}},\ \Eprint {https://arxiv.org/abs/1806.10820}
  {arXiv:1806.10820 [hep-ph]} \BibitemShut {NoStop}%
\bibitem [{pyt()}]{pythiaurl}%
  \BibitemOpen
  \href {https://pythia.org} {}\bibinfo {howpublished}
  {\url{https://pythia.org}}\BibitemShut {NoStop}%
\bibitem [{\citenamefont {Freer}\ \emph {et~al.}(2018)\citenamefont {Freer},
  \citenamefont {Horiuchi}, \citenamefont {Kanada-En'yo}, \citenamefont {Lee},\
  and\ \citenamefont {Mei{\ss}ner}}]{Freer:2017gip}%
  \BibitemOpen
  \bibfield  {author} {\bibinfo {author} {\bibfnamefont {M.}~\bibnamefont
  {Freer}}, \bibinfo {author} {\bibfnamefont {H.}~\bibnamefont {Horiuchi}},
  \bibinfo {author} {\bibfnamefont {Y.}~\bibnamefont {Kanada-En'yo}}, \bibinfo
  {author} {\bibfnamefont {D.}~\bibnamefont {Lee}},\ and\ \bibinfo {author}
  {\bibfnamefont {U.-G.}\ \bibnamefont {Mei{\ss}ner}},\ }\bibfield  {title}
  {\bibinfo {title} {{Microscopic Clustering in Light Nuclei}},\ }\href
  {https://doi.org/10.1103/RevModPhys.90.035004} {\bibfield  {journal}
  {\bibinfo  {journal} {Rev. Mod. Phys.}\ }\textbf {\bibinfo {volume} {90}},\
  \bibinfo {pages} {035004} (\bibinfo {year} {2018})},\ \Eprint
  {https://arxiv.org/abs/1705.06192} {arXiv:1705.06192 [nucl-th]} \BibitemShut
  {NoStop}%
\bibitem [{\citenamefont {Lee}(2009)}]{Lee:2008fa}%
  \BibitemOpen
  \bibfield  {author} {\bibinfo {author} {\bibfnamefont {D.}~\bibnamefont
  {Lee}},\ }\bibfield  {title} {\bibinfo {title} {{Lattice simulations for few-
  and many-body systems}},\ }\href {https://doi.org/10.1016/j.ppnp.2008.12.001}
  {\bibfield  {journal} {\bibinfo  {journal} {Prog. Part. Nucl. Phys.}\
  }\textbf {\bibinfo {volume} {63}},\ \bibinfo {pages} {117} (\bibinfo {year}
  {2009})},\ \Eprint {https://arxiv.org/abs/0804.3501} {arXiv:0804.3501
  [nucl-th]} \BibitemShut {NoStop}%
\bibitem [{\citenamefont {Giacalone}\ \emph
  {et~al.}(2025{\natexlab{b}})\citenamefont {Giacalone} \emph
  {et~al.}}]{Giacalone:2024luz}%
  \BibitemOpen
  \bibfield  {author} {\bibinfo {author} {\bibfnamefont {G.}~\bibnamefont
  {Giacalone}} \emph {et~al.},\ }\bibfield  {title} {\bibinfo {title}
  {{Exploiting Ne20 Isotopes for Precision Characterizations of Collectivity in
  Small Systems}},\ }\href {https://doi.org/10.1103/k8rb-jgvq} {\bibfield
  {journal} {\bibinfo  {journal} {Phys. Rev. Lett.}\ }\textbf {\bibinfo
  {volume} {135}},\ \bibinfo {pages} {012302} (\bibinfo {year}
  {2025}{\natexlab{b}})},\ \Eprint {https://arxiv.org/abs/2402.05995}
  {arXiv:2402.05995 [nucl-th]} \BibitemShut {NoStop}%
\bibitem [{\citenamefont {Oliinychenko}\ and\ \citenamefont
  {Petersen}(2016)}]{Oliinychenko:2015lva}%
  \BibitemOpen
  \bibfield  {author} {\bibinfo {author} {\bibfnamefont {D.}~\bibnamefont
  {Oliinychenko}}\ and\ \bibinfo {author} {\bibfnamefont {H.}~\bibnamefont
  {Petersen}},\ }\bibfield  {title} {\bibinfo {title} {{Deviations of the
  Energy-Momentum Tensor from Equilibrium in the Initial State for
  Hydrodynamics from Transport Approaches}},\ }\href
  {https://doi.org/10.1103/PhysRevC.93.034905} {\bibfield  {journal} {\bibinfo
  {journal} {Phys. Rev. C}\ }\textbf {\bibinfo {volume} {93}},\ \bibinfo
  {pages} {034905} (\bibinfo {year} {2016})},\ \Eprint
  {https://arxiv.org/abs/1508.04378} {arXiv:1508.04378 [nucl-th]} \BibitemShut
  {NoStop}%
\bibitem [{\citenamefont {Inghirami}\ and\ \citenamefont
  {Elfner}(2022)}]{Inghirami:2022afu}%
  \BibitemOpen
  \bibfield  {author} {\bibinfo {author} {\bibfnamefont {G.}~\bibnamefont
  {Inghirami}}\ and\ \bibinfo {author} {\bibfnamefont {H.}~\bibnamefont
  {Elfner}},\ }\bibfield  {title} {\bibinfo {title} {{The applicability of
  hydrodynamics in heavy ion collisions at
  $\sqrt{s_\mathrm{NN}}$~=~2.4{\textendash}7.7~GeV}},\ }\href
  {https://doi.org/10.1140/epjc/s10052-022-10718-x} {\bibfield  {journal}
  {\bibinfo  {journal} {Eur. Phys. J. C}\ }\textbf {\bibinfo {volume} {82}},\
  \bibinfo {pages} {796} (\bibinfo {year} {2022})},\ \Eprint
  {https://arxiv.org/abs/2201.05934} {arXiv:2201.05934 [hep-ph]} \BibitemShut
  {NoStop}%
\bibitem [{\citenamefont {Noronha}\ \emph {et~al.}(2024)\citenamefont
  {Noronha}, \citenamefont {Schenke}, \citenamefont {Shen},\ and\ \citenamefont
  {Zhao}}]{Noronha:2024dtq}%
  \BibitemOpen
  \bibfield  {author} {\bibinfo {author} {\bibfnamefont {J.}~\bibnamefont
  {Noronha}}, \bibinfo {author} {\bibfnamefont {B.}~\bibnamefont {Schenke}},
  \bibinfo {author} {\bibfnamefont {C.}~\bibnamefont {Shen}},\ and\ \bibinfo
  {author} {\bibfnamefont {W.}~\bibnamefont {Zhao}},\ }\bibfield  {title}
  {\bibinfo {title} {{Progress and challenges in small systems}},\ }\href
  {https://doi.org/10.1142/9789811294679_0004} {\bibfield  {journal} {\bibinfo
  {journal} {Int. J. Mod. Phys. E}\ }\textbf {\bibinfo {volume} {33}},\
  \bibinfo {pages} {2430005} (\bibinfo {year} {2024})},\ \Eprint
  {https://arxiv.org/abs/2401.09208} {arXiv:2401.09208 [nucl-th]} \BibitemShut
  {NoStop}%
\bibitem [{\citenamefont {Kurkela}\ \emph {et~al.}(2019)\citenamefont
  {Kurkela}, \citenamefont {Wiedemann},\ and\ \citenamefont
  {Wu}}]{Kurkela:2019kip}%
  \BibitemOpen
  \bibfield  {author} {\bibinfo {author} {\bibfnamefont {A.}~\bibnamefont
  {Kurkela}}, \bibinfo {author} {\bibfnamefont {U.~A.}\ \bibnamefont
  {Wiedemann}},\ and\ \bibinfo {author} {\bibfnamefont {B.}~\bibnamefont
  {Wu}},\ }\bibfield  {title} {\bibinfo {title} {{Flow in AA and pA as an
  interplay of fluid-like and non-fluid like excitations}},\ }\href
  {https://doi.org/10.1140/epjc/s10052-019-7428-6} {\bibfield  {journal}
  {\bibinfo  {journal} {Eur. Phys. J. C}\ }\textbf {\bibinfo {volume} {79}},\
  \bibinfo {pages} {965} (\bibinfo {year} {2019})},\ \Eprint
  {https://arxiv.org/abs/1905.05139} {arXiv:1905.05139 [hep-ph]} \BibitemShut
  {NoStop}%
\bibitem [{\citenamefont {Ambrus}\ \emph
  {et~al.}(2023{\natexlab{a}})\citenamefont {Ambrus}, \citenamefont
  {Schlichting},\ and\ \citenamefont {Werthmann}}]{Ambrus:2022koq}%
  \BibitemOpen
  \bibfield  {author} {\bibinfo {author} {\bibfnamefont {V.~E.}\ \bibnamefont
  {Ambrus}}, \bibinfo {author} {\bibfnamefont {S.}~\bibnamefont
  {Schlichting}},\ and\ \bibinfo {author} {\bibfnamefont {C.}~\bibnamefont
  {Werthmann}},\ }\bibfield  {title} {\bibinfo {title} {{Opacity dependence of
  transverse flow, preequilibrium, and applicability of hydrodynamics in
  heavy-ion collisions}},\ }\href {https://doi.org/10.1103/PhysRevD.107.094013}
  {\bibfield  {journal} {\bibinfo  {journal} {Phys. Rev. D}\ }\textbf {\bibinfo
  {volume} {107}},\ \bibinfo {pages} {094013} (\bibinfo {year}
  {2023}{\natexlab{a}})},\ \Eprint {https://arxiv.org/abs/2211.14379}
  {arXiv:2211.14379 [hep-ph]} \BibitemShut {NoStop}%
\bibitem [{\citenamefont {Ambrus}\ \emph
  {et~al.}(2023{\natexlab{b}})\citenamefont {Ambrus}, \citenamefont
  {Schlichting},\ and\ \citenamefont {Werthmann}}]{Ambrus:2022qya}%
  \BibitemOpen
  \bibfield  {author} {\bibinfo {author} {\bibfnamefont {V.~E.}\ \bibnamefont
  {Ambrus}}, \bibinfo {author} {\bibfnamefont {S.}~\bibnamefont
  {Schlichting}},\ and\ \bibinfo {author} {\bibfnamefont {C.}~\bibnamefont
  {Werthmann}},\ }\bibfield  {title} {\bibinfo {title} {{Establishing the Range
  of Applicability of Hydrodynamics in High-Energy Collisions}},\ }\href
  {https://doi.org/10.1103/PhysRevLett.130.152301} {\bibfield  {journal}
  {\bibinfo  {journal} {Phys. Rev. Lett.}\ }\textbf {\bibinfo {volume} {130}},\
  \bibinfo {pages} {152301} (\bibinfo {year} {2023}{\natexlab{b}})},\ \Eprint
  {https://arxiv.org/abs/2211.14356} {arXiv:2211.14356 [hep-ph]} \BibitemShut
  {NoStop}%
\bibitem [{\citenamefont {Borghini}\ \emph {et~al.}(2001)\citenamefont
  {Borghini}, \citenamefont {Dinh},\ and\ \citenamefont
  {Ollitrault}}]{Borghini:2000sa}%
  \BibitemOpen
  \bibfield  {author} {\bibinfo {author} {\bibfnamefont {N.}~\bibnamefont
  {Borghini}}, \bibinfo {author} {\bibfnamefont {P.~M.}\ \bibnamefont {Dinh}},\
  and\ \bibinfo {author} {\bibfnamefont {J.-Y.}\ \bibnamefont {Ollitrault}},\
  }\bibfield  {title} {\bibinfo {title} {{A New method for measuring azimuthal
  distributions in nucleus-nucleus collisions}},\ }\href
  {https://doi.org/10.1103/PhysRevC.63.054906} {\bibfield  {journal} {\bibinfo
  {journal} {Phys. Rev. C}\ }\textbf {\bibinfo {volume} {63}},\ \bibinfo
  {pages} {054906} (\bibinfo {year} {2001})},\ \Eprint
  {https://arxiv.org/abs/nucl-th/0007063} {arXiv:nucl-th/0007063} \BibitemShut
  {NoStop}%
\bibitem [{\citenamefont {Bilandzic}(2012)}]{Bilandzic:2012wva}%
  \BibitemOpen
  \bibfield  {author} {\bibinfo {author} {\bibfnamefont {A.}~\bibnamefont
  {Bilandzic}},\ }\emph {\bibinfo {title} {{Anisotropic flow measurements in
  ALICE at the large hadron collider}}},\ \href@noop {} {Ph.D. thesis},\
  \bibinfo  {school} {Utrecht U.} (\bibinfo {year} {2012})\BibitemShut
  {NoStop}%
\bibitem [{\citenamefont {Bilandzic}\ \emph {et~al.}(2011)\citenamefont
  {Bilandzic}, \citenamefont {Snellings},\ and\ \citenamefont
  {Voloshin}}]{Bilandzic:2010jr}%
  \BibitemOpen
  \bibfield  {author} {\bibinfo {author} {\bibfnamefont {A.}~\bibnamefont
  {Bilandzic}}, \bibinfo {author} {\bibfnamefont {R.}~\bibnamefont
  {Snellings}},\ and\ \bibinfo {author} {\bibfnamefont {S.}~\bibnamefont
  {Voloshin}},\ }\bibfield  {title} {\bibinfo {title} {{Flow analysis with
  cumulants: Direct calculations}},\ }\href
  {https://doi.org/10.1103/PhysRevC.83.044913} {\bibfield  {journal} {\bibinfo
  {journal} {Phys. Rev. C}\ }\textbf {\bibinfo {volume} {83}},\ \bibinfo
  {pages} {044913} (\bibinfo {year} {2011})},\ \Eprint
  {https://arxiv.org/abs/1010.0233} {arXiv:1010.0233 [nucl-ex]} \BibitemShut
  {NoStop}%
\bibitem [{\citenamefont {Jia}\ \emph {et~al.}(2017)\citenamefont {Jia},
  \citenamefont {Zhou},\ and\ \citenamefont {Trzupek}}]{Jia:2017hbm}%
  \BibitemOpen
  \bibfield  {author} {\bibinfo {author} {\bibfnamefont {J.}~\bibnamefont
  {Jia}}, \bibinfo {author} {\bibfnamefont {M.}~\bibnamefont {Zhou}},\ and\
  \bibinfo {author} {\bibfnamefont {A.}~\bibnamefont {Trzupek}},\ }\bibfield
  {title} {\bibinfo {title} {{Revealing long-range multiparticle collectivity
  in small collision systems via subevent cumulants}},\ }\href
  {https://doi.org/10.1103/PhysRevC.96.034906} {\bibfield  {journal} {\bibinfo
  {journal} {Phys. Rev. C}\ }\textbf {\bibinfo {volume} {96}},\ \bibinfo
  {pages} {034906} (\bibinfo {year} {2017})},\ \Eprint
  {https://arxiv.org/abs/1701.03830} {arXiv:1701.03830 [nucl-th]} \BibitemShut
  {NoStop}%
\bibitem [{\citenamefont {Song}\ and\ \citenamefont
  {Heinz}(2008)}]{Song:2008si}%
  \BibitemOpen
  \bibfield  {author} {\bibinfo {author} {\bibfnamefont {H.}~\bibnamefont
  {Song}}\ and\ \bibinfo {author} {\bibfnamefont {U.~W.}\ \bibnamefont
  {Heinz}},\ }\bibfield  {title} {\bibinfo {title} {{Multiplicity scaling in
  ideal and viscous hydrodynamics}},\ }\href
  {https://doi.org/10.1103/PhysRevC.78.024902} {\bibfield  {journal} {\bibinfo
  {journal} {Phys. Rev. C}\ }\textbf {\bibinfo {volume} {78}},\ \bibinfo
  {pages} {024902} (\bibinfo {year} {2008})},\ \Eprint
  {https://arxiv.org/abs/0805.1756} {arXiv:0805.1756 [nucl-th]} \BibitemShut
  {NoStop}%
\bibitem [{\citenamefont {Saraswat}\ \emph {et~al.}(2018)\citenamefont
  {Saraswat}, \citenamefont {Shukla},\ and\ \citenamefont
  {Singh}}]{Saraswat:2017kpg}%
  \BibitemOpen
  \bibfield  {author} {\bibinfo {author} {\bibfnamefont {K.}~\bibnamefont
  {Saraswat}}, \bibinfo {author} {\bibfnamefont {P.}~\bibnamefont {Shukla}},\
  and\ \bibinfo {author} {\bibfnamefont {V.}~\bibnamefont {Singh}},\ }\bibfield
   {title} {\bibinfo {title} {{Transverse momentum spectra of hadrons in high
  energy pp and heavy ion collisions}},\ }\href
  {https://doi.org/10.1088/2399-6528/aab00f} {\bibfield  {journal} {\bibinfo
  {journal} {J. Phys. Comm.}\ }\textbf {\bibinfo {volume} {2}},\ \bibinfo
  {pages} {035003} (\bibinfo {year} {2018})},\ \Eprint
  {https://arxiv.org/abs/1706.04860} {arXiv:1706.04860 [hep-ph]} \BibitemShut
  {NoStop}%
\bibitem [{\citenamefont {Braun-Munzinger}\ \emph {et~al.}(1995)\citenamefont
  {Braun-Munzinger}, \citenamefont {Stachel}, \citenamefont {Wessels},\ and\
  \citenamefont {Xu}}]{Braun-Munzinger:1994ewq}%
  \BibitemOpen
  \bibfield  {author} {\bibinfo {author} {\bibfnamefont {P.}~\bibnamefont
  {Braun-Munzinger}}, \bibinfo {author} {\bibfnamefont {J.}~\bibnamefont
  {Stachel}}, \bibinfo {author} {\bibfnamefont {J.~P.}\ \bibnamefont
  {Wessels}},\ and\ \bibinfo {author} {\bibfnamefont {N.}~\bibnamefont {Xu}},\
  }\bibfield  {title} {\bibinfo {title} {{Thermal equilibration and expansion
  in nucleus-nucleus collisions at the AGS}},\ }\href
  {https://doi.org/10.1016/0370-2693(94)01534-J} {\bibfield  {journal}
  {\bibinfo  {journal} {Phys. Lett. B}\ }\textbf {\bibinfo {volume} {344}},\
  \bibinfo {pages} {43} (\bibinfo {year} {1995})},\ \Eprint
  {https://arxiv.org/abs/nucl-th/9410026} {arXiv:nucl-th/9410026} \BibitemShut
  {NoStop}%
\bibitem [{\citenamefont {Ollitrault}\ \emph {et~al.}(2009)\citenamefont
  {Ollitrault}, \citenamefont {Poskanzer},\ and\ \citenamefont
  {Voloshin}}]{Ollitrault:2009ie}%
  \BibitemOpen
  \bibfield  {author} {\bibinfo {author} {\bibfnamefont {J.-Y.}\ \bibnamefont
  {Ollitrault}}, \bibinfo {author} {\bibfnamefont {A.~M.}\ \bibnamefont
  {Poskanzer}},\ and\ \bibinfo {author} {\bibfnamefont {S.~A.}\ \bibnamefont
  {Voloshin}},\ }\bibfield  {title} {\bibinfo {title} {{Effect of flow
  fluctuations and nonflow on elliptic flow methods}},\ }\href
  {https://doi.org/10.1103/PhysRevC.80.014904} {\bibfield  {journal} {\bibinfo
  {journal} {Phys. Rev. C}\ }\textbf {\bibinfo {volume} {80}},\ \bibinfo
  {pages} {014904} (\bibinfo {year} {2009})},\ \Eprint
  {https://arxiv.org/abs/0904.2315} {arXiv:0904.2315 [nucl-ex]} \BibitemShut
  {NoStop}%
\bibitem [{\citenamefont {Aad}\ \emph {et~al.}(2025)\citenamefont {Aad} \emph
  {et~al.}}]{ATLAS:2025nnt}%
  \BibitemOpen
  \bibfield  {author} {\bibinfo {author} {\bibfnamefont {G.}~\bibnamefont
  {Aad}} \emph {et~al.} (\bibinfo {collaboration} {ATLAS}),\ }\bibfield
  {title} {\bibinfo {title} {{Measurement of the azimuthal anisotropy of
  charged particles in $\sqrt{s_{\mathrm{NN}}}=5.36$ TeV $^{16}$O$+^{16}$O and
  $^{20}$Ne$+^{20}$Ne collisions with the ATLAS detector}},\ }\href@noop {} {\
  (\bibinfo {year} {2025})},\ \Eprint {https://arxiv.org/abs/2509.05171}
  {arXiv:2509.05171 [nucl-ex]} \BibitemShut {NoStop}%
\end{thebibliography}%
\end{document}